\documentclass[12pt]{article}
\pdfoutput=1

\usepackage{fullpage}
\usepackage{graphicx}
\usepackage{cite}
\usepackage{authblk}
\usepackage{makecell,booktabs,array,subcaption}
\newcolumntype{x}[1]{%
>{\centering\hspace{0pt}}p{#1}}%
\newcommand{\tnhl}{\tabularnewline\hline}
\newcommand{\tn}{\tabularnewline}

\usepackage{amsmath}
\usepackage{amssymb}
\usepackage{subcaption}
\usepackage[htt]{hyphenat}

\newcommand{\msbar}{\overline{\text{MS}}}

\usepackage{graphicx}
\usepackage{dcolumn}
\usepackage{bm}
\usepackage{color,soul}

\usepackage{pgfplots}
\usepackage{tikz}
\usepackage{hyperref}

\pgfplotsset{compat=1.14}

\definecolor{darkyellow}{rgb}{0.5, 0.5, 0.0}
\definecolor{darkpurple}{rgb}{0.5, 0.2, 0.8}
\definecolor{darkblue}{rgb}{0.0, 0.0, 0.8}
\definecolor{darkgreen}{rgb}{0.0, 0.4, 0.0}
\definecolor{darkred}{rgb}{0.5, 0.0, 0.0}
\usepackage{hyperref}
\hypersetup{
    linktocpage,
     colorlinks,
     citecolor=darkgreen,
     linkcolor= darkgreen,
     urlcolor=darkgreen
}

\newcommand{\gev}{\ensuremath{~\rm{GeV}}}
\newcommand{\mtmc}{m_t^{\text{MC}}}

\title{Parameter Inference from Event Ensembles and the Top-Quark Mass}

\author{Forrest Flesher}
\author{Katherine Fraser}
\author{Charles Hutchison}
\author{Bryan Ostdiek}
\author{Matthew D. Schwartz}

\affil{\small \emph{Department of Physics, Harvard University, Cambridge, MA 02138, USA}\\
\emph{The NSF AI Institute for Artificial Intelligence and Fundamental Interactions\\
~\\
\emph{E-mail:} 
forrestflesher@college.harvard.edu,
kfraser@g.harvard.edu,
hutchison@college.harvard.edu,
bostdiek@g.harvard.edu,
schwartz@g.harvard.edu
}}
\begin{document}

\unitlength = 0.4mm
\maketitle
\thispagestyle{empty}

\begin{abstract}
One of the key tasks of any particle collider is measurement. In practice, this is often done by fitting data to a simulation, which depends on many parameters. Sometimes, when the effects of varying different parameters are highly correlated, a large ensemble of data may be needed to resolve parameter-space degeneracies. 
An important example is measuring the top-quark mass, where  other physical and unphysical parameters in the simulation must be profiled when fitting the top-quark mass parameter. 
We compare four different methodologies for top-quark mass measurement: a classical histogram fit similar to one commonly used in experiment augmented by soft-drop jet grooming; a 2D profile likelihood fit with a nuisance parameter; a machine-learning method called DCTR; and a linear regression approach, either using a least-squares fit or with a dense linearly-activated neural network. Despite the fact that individual events are
totally uncorrelated, we find that the linear regression methods work most
effectively when we input an ensemble of events sorted by mass,
rather than training them on individual events. Although all methods provide robust extraction of the top-quark mass parameter, the linear network does marginally best and is remarkably simple. For the top study, we conclude that the Monte-Carlo-based uncertainty on current extractions of the top-quark mass from LHC data can be reduced significantly (by perhaps a factor of 2) using networks trained on sorted event ensembles. 
More generally, machine learning from ensembles for parameter estimation has broad potential for collider physics measurements. 
\end{abstract}

\newpage
\hypersetup{pageanchor=true}
\pagenumbering{arabic}

\maketitle

\section{Introduction}
The number one goal of collider physics experiments is to determine the existence and properties of particles in nature. In some rare cases first-principles theoretical calculations can be compared directly to data.
More commonly, theory is used to construct sophisticated simulations with adjustable parameters that are then fit to data. Some of these simulation parameters, like coupling constants or masses, have straightforward physical interpretations while other parameters, such as elements of Pythia's string fragmentation model~\cite{Sjostrand:2014zea}, are required to provide enough flexibility for the data to be described. Often the parameters are highly correlated: varying one can sometimes be entirely compensated by varying another. Typically the uncertainty generated by profiling the unphysical parameters is smaller than other sources of uncertainty, however for precision studies it can be important.

The example of parameter extraction studied in this paper is the determination of the top-quark mass. The top mass is one of the few parameters in the Standard Model for which a measurement with improved precision is both extremely important and feasible at the LHC. For example, our current best estimate of the lifetime of our metastable vacuum in the universe is limited by precision on top quark mass and the strong coupling constant~\cite{Coleman:1973jx,Isidori:2001bm,Degrassi:2012ry,Andreassen:2017rzq}. Moreover, the lifetime is exponentially sensitive to the top quark mass. Using $m_t^\text{pole} = 173.1$ GeV, our universe is predicted to last $10^{167}$ years, but if the top mass were 0.6 GeV higher it would last only $10^{111}$ years, and if it were 0.6 GeV lower, the universe would last $10^{252}$ years~\cite{Andreassen:2017rzq}.  
Another example, is searches for certain supersymmetry (SUSY) models, in which stop squarks that are nearly degenerate with the top quark are difficult to constrain because the signal is so similar to $t\bar{t}$ background. This similarity allows the stops to contaminate precision measurements of the top quark, so the consistency of top measurements can be used to search the SUSY parameter space~\cite{Czakon:2014fka,Eifert:2014kea,Aad:2014kva,Cohen:2018arg,Cohen:2019ycc,Aaboud:2019hwz}.

It is possible to measure the top-quark mass by direct theory-data comparison, for example through total cross section measurements~\cite{cms2019measurement}. The cross-section approach has two main advantages: it allows for a direct comparison between data and precision theory and the top mass extracted has a clean short-distance definition (typically the $\msbar$ mass).
However, current mass determination by this method has an uncertainty of 1-2 GeV~\cite{Aad:2014kva,Khachatryan:2016mqs,Sirunyan:2018goh,ATLAS:2019uka,Aaboud:2017ujq,Sirunyan:2019zvx}.
The method for extracting the top-quark mass from LHC data that currently has the smallest uncertainty is fitting the invariant mass peak from the decay products of top quarks in $t\bar{t}$ events~\cite{Aaboud:2016igd,Aaboud:2017mae,Aaboud:2018zbu,Sirunyan:2018gqx,Sirunyan:2018mlv}. 
While such fits typically have errors at the sub-GeV level, there are systematic and theoretical uncertainties associated with such a procedure that are not present in the cross section method.
The main complication is that one is more reliant on simulation.
For example, there is an uncertainty about how to translate the mass extracted this way, called the Monte Carlo mass, to a scheme like the $\msbar$ mass which is more theoretically sound. 

It is important to separate the challenges in converting between a Monte Carlo mass parameter and a short distance scheme like $\msbar$ from the extraction of the Monte Carlo mass parameter itself. Typically, the conversion to $\msbar$ is done by equating the Monte Carlo mass with the pole mass. One could attempt to systematically improve this mapping, for example by comparing precision theory and simulation directly (without data)~\cite{Hoang:2018zrp,Fleming:2007qr,Hoang:2008xm,Butenschoen:2016lpz,Hoang:2017kmk,Hoang:2020iah,PhysRevLett.116.162001}.
Regardless of how or whether this is done, one cannot hope to begin converting from the Monte Carlo mass to another scheme if different Monte Carlo tunes lead to a different value of the top mass when fit to the same data. Thus, a prerequisite for considering the conversion between Monte Carlo and pole mass is to reduce the tune-dependence of the extracted mass. This reduction is the primary target of this paper.

The problem of reducing tune uncertainty of the top mass was examined in~\cite{Andreassen:2017ugs}. There it was estimated that using classical histogram fitting, the tune-uncertainty on the top Monte Carlo mass was around 500 MeV. This number results from a comparison among the masses extracted using a standard set of tunes in Pythia. It was then shown that the uncertainty could be reduced to 200 MeV by calibrating to the $W$ mass (as is often done by the experiments), and further reduced to 140 MeV by applying soft-drop jet grooming~\cite{Larkoski:2014wba} to the data before fitting. In this paper, we reproduce the main results of~\cite{Andreassen:2017ugs} and explore whether further reduction is possible using machine learning or with linear regression on ensembles of events. Additionally, we also compare these methods to a profile likelihood fit that is similar to what is currently done in the best experimental measurements.

Using machine learning (ML) to fit a parameter like the top Monte Carlo mass involves complementary challenges to typical collider physics ML applications. Typical collider ML applications such as top-tagging or pileup removal have essentially a right answer: which event was a top and which was background, or what does an event look like with pileup removed? For top mass measurement, there is no right answer: a perfect oracle would not be able to determine the mass from a single event. Instead, only after a collection of events are observed can the top mass be extracted.

There are a number of approaches that have been suggested for learning from ensembles of events. For example, the JUNIPR framework uses a jet-physics inspired architecture to construct the likelihood distribution~\cite{Andreassen:2018apy,Andreassen:2019txo}. This can be done as a function of the top mass, or other training parameters which can then be regressed on data. This application for JUNIPR was suggested in~\cite{Andreassen:2019txo} but has not yet been implemented to our knowledge.

\begin{figure}[t]
    \centering
    \includegraphics[width=0.65\linewidth]{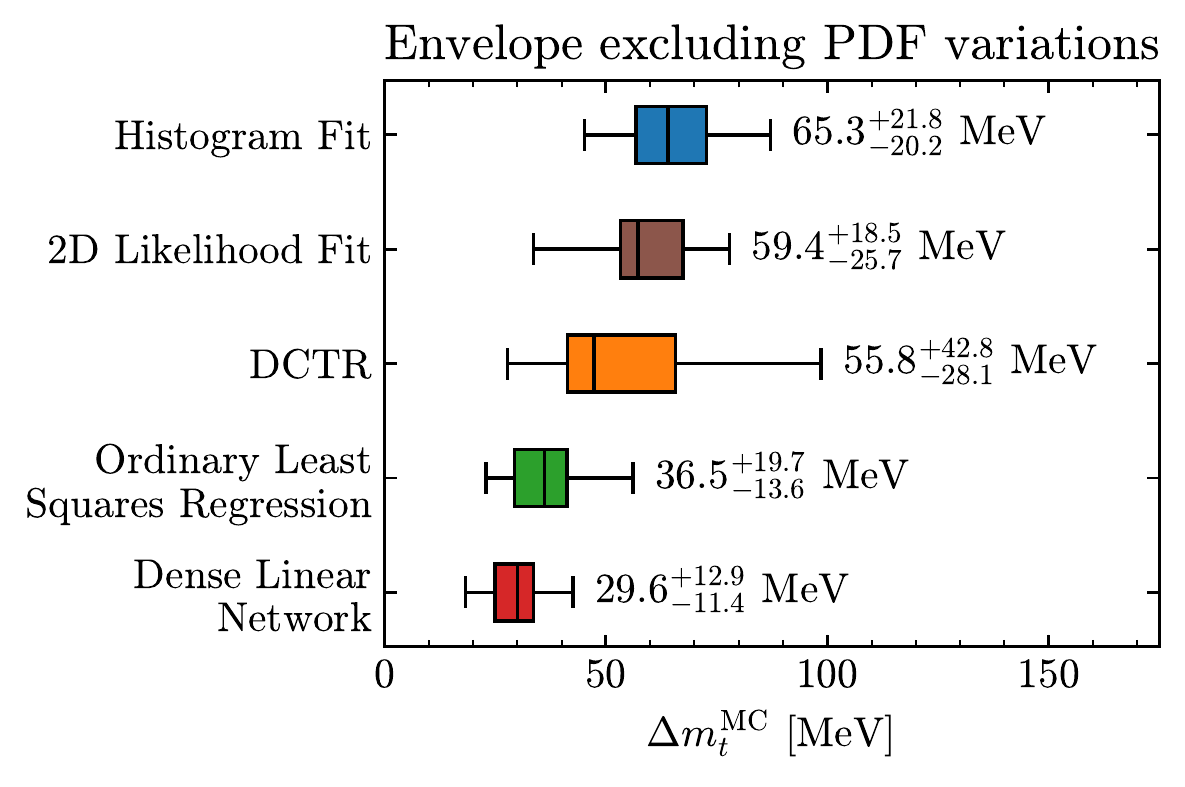}
    \caption{Summary of main results. The bars show the estimate of the Monte-Carlo tuning uncertainty in top-quark mass extraction from top events at the LHC. The errors on the uncertainties include  uncertainties from training and statistical variations.
    The top row is a histogram fit, using soft drop and normalizing to the $W$-mass (following~\cite{Andreassen:2017ugs}).
    The second row uses a 2D likelihood fit to profile over correlations between the top and $W$ masses.
    The third row uses the DCTR technique of~\cite{Andreassen:2019nnm}.
    The fourth row is an ordinary least squares linear regression on an ensemble of 30,000 events.
    The fifth row shows the result of using a linear network for regression, also on an ensemble of 30,000 events. 
    Numbers here correspond to the total envelope excluding PDF uncertainties, as in Fig.~\ref{fig:AllMethods}.
    }
    \label{fig:ErrorSummary}
\end{figure}

Another approach is the DCTR method proposed in~\cite{Andreassen:2019nnm}. DCTR works by learning the relative weights of a distribution of events as a function of some reference tuning parameters. Then it can be used for regression by minimizing the loss over the tuning parameters to find the best fit.
DCTR takes as input events processed through a Particle Flow Network\cite{Komiske:2018cqr}, which is an adaptation of the ``Deep Sets'' framework developed in~\cite{NIPS2017} to particle physics. In~\cite{Andreassen:2019nnm}, it was shown to be able to fit simultaneously three Monte Carlo tuning parameters in $e^+e^- \to $ jet events with good results. Thus it is a natural candidate method to test on top-mass extraction where there is a clear metric for what a ``good" fit would be. Although the top mass is physical, the top mass parameter in the Monte Carlo can be treated as a tuning parameter and fit in the same way as other Monte Carlo parameters.  A discussion of the DCTR method is given in Section~\ref{sec:DCTR}.

While the DCTR method is promising, it is somewhat cumbersome to implement and train.
Moreover, learning the full likelihood ratio as a function of a very high-dimensional input (such as ParticleFlow) may not be necessary if the goal is the regression of a single parameter, like the top mass.
Thus we also consider a simpler approach, where ordered sets of high-level observables are input to a dense neural network. We discuss this approach in Section~\ref{sec:dense}. The dense network is very effective, even if the activation functions connecting the nodes are linear. Thus, the entire network is a linear function acting on a sorted ensemble of events. We compare the linear network performance to an ordinary least squares regression, finding similar performance. Moreover, the linear mapping can be examined to see how it depends on the tune and the various elements of the input ensemble. This analysis is also included in Section~\ref{sec:dense}.
A summary of our main findings is given in Fig.~\ref{fig:ErrorSummary}.

The paper is organized as follows. Event generation and general elements of our fitting procedure are discussed in Section~\ref{sec:evtgeneration}. Section~\ref{sec:Classical} describes our implementation of classical fitting methods that are similar to what is often done in experimental work, including a histogram fit modeled after~\cite{Andreassen:2017ugs} and a 2D profile likelihood fit, to benchmark our samples and fits. Section~\ref{sec:dense} discusses the regression approach, using both a linear network and an ordinary least squares regression. Section~\ref{sec:DCTR} discusses the DCTR approach of~\cite{Andreassen:2019nnm}. Our conclusions and a discussion are given in Section~\ref{sec:discussion}.

\section{Event Generation and Uncertainty Estimation}
\label{sec:evtgeneration}
For this study, events with pairs of top quarks are produced using \textsc{Pythia 8}, including both $q\bar{q}$ and $gg$ production channels in $\sqrt{s}=13$ TeV proton-proton collisions.
We restrict to semi-leptonic events with $t\rightarrow b ~\ell^+ ~\nu_{\ell}$ and $\bar{t} \rightarrow \bar{b} ~q~\bar{q}^{\prime}$, where $\ell$ stands for electrons or muons. 
The events are showered to final state particles, which are then clustered into jets using the anti-$k_t$~\cite{Cacciari:2008gp} algorithm in \textsc{Fastjet}~\cite{Cacciari:2011ma} with R=0.5.
For simplicity, we do not attempt to include any realistic detector effects or experimental efficiencies. Thus, we mark any jet within $\Delta R < 0.4$ of one of the $b$'s from the top decays as a $b$-jet and assume the lepton is correctly tagged.
A more realistic study would of course need to incorporate $b$-tagging, jet energy scale resolution, pileup, backgrounds, and so on. Each of these effects will necessarily increase the top-quark mass uncertainty. However, since our goal is mainly to understand the relative performance of different ensemble regression methods, we do not believe our simplifying assumptions will affect the qualitative conclusions.

The event selection is as follows. We require a final state $\ell$ with $p_T^{\ell} > 20\gev$  and $\big|\eta \big| < 2.4$. We only keep jets if they have $p_T^j > 30\gev$ and $\big|\eta\big| < 2.4$ and demand that there are exactly 2 $b$-tagged jets and at least 2 un-tagged jets.
The invariant mass of pairs of un-tagged jets is scanned to find the pair with a mass closest to $m_W = 80.3\gev$.
If this two-jet invariant mass, $m_{2j}$ is not within $(70\gev, 90\gev)$, the event is discarded.
Next, we find the three-jet invariant mass for the two jets of the $W$ and $b$-tagged jet coming from the $\bar{b}$.
This is overly simplified and ignores combinatoric background.\footnote{Ref.~\cite{Fenton:2020woz} introduces a machine learning method to identify the correct combination of jets in $t\bar{t}$ without the factorial scaling of scanning each combinatorial permutation.}
However, we take a tight cut on the three-jet invariant mass, and only accept events with $150\gev < m_{3j} < 200\gev$\footnote{In the dense network section, we also generate events without this cut to see the effect of the $m_{3j}$ range on $\Delta \mtmc$.}, which reduces such contamination.
This still allows for a comparison of the different methods.

The uncertainty in the regression of the Monte Carlo top mass from each of the methods is computed using the A14 Pythia 8 Tunes of the ATLAS 7 TeV data~\cite{TheATLAScollaboration:2014rfk}. The 14 tunes cover 4 different families of variations: VarPDF, Var1, Var2, and Var3. VarPDF covers variations in the parton distribution functions with \texttt{tunepp:19-22} corresponding to the CTEQL1~\cite{Pumplin:2002vw}, MSTW2008LO~\cite{Watt:2012tq}, NNPDF2.3LO~\cite{Carrazza:2013axa}, and HERAPDF1.5LO~\cite{Sarkar:2014zua} PDFs, respectively. Var1, Var2, and Var3 all use the NNPDF2.3LO~\cite{Carrazza:2013axa} PDF but vary other parameters, with Var1 covering underlying event effects, Var2 accounting for jet substructure, and Var3 covering different aspects of extra jet production. Var3 includes three separate variations (Var3a, Var3b, and Var3c) since it could not be reduced to a single pair. The tuning parameters for all A14 variations are shown in table~\ref{label:tuning_parmas}.

There are of course many more tunes one can consider. But again, since the main purpose of this study is to compare the relative strengths of different approaches, not to produce a final numerical value of the uncertainty, we believe this set should be sufficient.

\begin{table}[t!]\centering
\begin{tabular}{p{2.5cm}|x{2.5cm}|x{2.5cm}|x{2.5cm}|x{2.5cm}}
\toprule
 Variation & Tunes & ColorRec & $\alpha_S^{\text{MPI}}$ & $p_{T0}^\text{Ref}$  \tnhl
 \hline
 VarPDF & 19-22 & 1.71 & 0.126 & 1.56  \tnhl
 Var1 & 21, 23, 24 & [1.69,1.73] & [0.121,0.131] & 1.56  \tnhl
 Var2 & 21, 25, 26  & 1.71 & 0.126 & [1.50,1.60] \tnhl
 Var3a  & 21, 27, 28 & 1.71 & [0.125,0.127] &  [1.51,1.67]  \tnhl
 Var3b  & 21, 29, 30 & 1.71 & 0.126 &  1.56  \tnhl
 Var3c  & 21, 31, 32 & 1.71 & 0.126 & 1.56  \tn
 \bottomrule
\end{tabular}
\begin{tabular}{p{2.5cm}|x{2.5cm}|x{2.5cm}|x{2.5cm}|x{2.5cm}}
\toprule
 Variation & $p_T^\text{dampFudge}$ &  $\alpha_S^{\text{FSR}}$ & $p_T^\text{maxFudge}$ & $\alpha_S^{\text{ISR}}$ \tnhl \hline
 VarPDF  & 1.05 & 0.127 & 0.91 & 0.127 \tnhl
 Var1 & 1.05  & 0.127 & 0.91 & 0.127 \tnhl
 Var2 & [1.04,1.08] & [0.124,0.136] & 0.91 & 0.127  \tnhl
 Var3a & [0.93,1.36] & [0.124,0.136] & [0.88,0.98] & 0.127 \tnhl
 Var3b & [1.04,1.07] & [0.114,0.138] & [0.83,1.00] & [0.126,0.129] \tnhl
 Var3c & 1.05 & 0.127 & 0.91 & [0.115,0.140] \tn
 \bottomrule
\end{tabular}
\caption{Table shows the relevant parameters for the A14 tune variations. Var1 through Var3 tunes are listed in order of central, +, then -. The relevant tuning parameters are the strength of the color reconnection (\texttt{ColourReconnection:range}), the strong coupling constant for multiparticle interactions $\alpha_S^{MPI}$ (\texttt{MultipartonInteractions:alphaSvalue}), the initial state radiation (ISR) $p_T$ cutoff $p_{T0}^\text{Ref}$  (\texttt{SpaceShower:pT0Ref}), the factorization/renormalization scale damping $p_T^\text{dampFudge}$ (\texttt{SpaceShower:pTdampFudge}), the strong coupling constant for final state radiation (FSR) $\alpha_S^{\text{FSR}}$ (\texttt{TimeShower:alphaSvalue}), the multiplicative factor on the max ISR evolution scale $p_T^\text{maxFudge}$ (\texttt{SpaceShower:pTmaxFudge}), and the ISR strong coupling  $\alpha_{S}^{\text{ISR}}$ (\texttt{SpaceShower:alphaSvalue}).}
\label{label:tuning_parmas}
\end{table}

We attempt as much as is possible to use the same fitting procedure to compare different methods. In all cases, after a method is fit or trained, it provides a mapping from an ensemble of events to a regressed mass. To assess the uncertainty of the regressed mass, we first assess its variation for fixed Monte Carlo mass within each tune family.
We denote the maximum, minimum and mean regressed mass within the family for the fixed mass by $m_{\rm{fit}}^{\rm{max}}$, $m_{\rm{fit}}^{\rm{min}}$, and $\bar{m}_{\rm{fit}}$, respectively.
We compute the uncertainty for the given $\mtmc$ and tune family as
\begin{equation}
    \Delta \mtmc = \frac{1}{2} \big(m_{\rm{fit}}^{\rm{max}} - m_{\rm{fit}}^{\rm{min}} \big) \frac{\mtmc}{\bar{m}_t^{\text{fit}}}.
\label{eqn:error estimate}
\end{equation}
The factor of $\mtmc / \bar{m}_t^{\text{fit}}$ reflects that the fit mass (especially in the histogram fit with soft drop) can be linearly offset from the true Monte Carlo mass.

\begin{figure}[t]
    \centering
    \includegraphics[width=\linewidth]{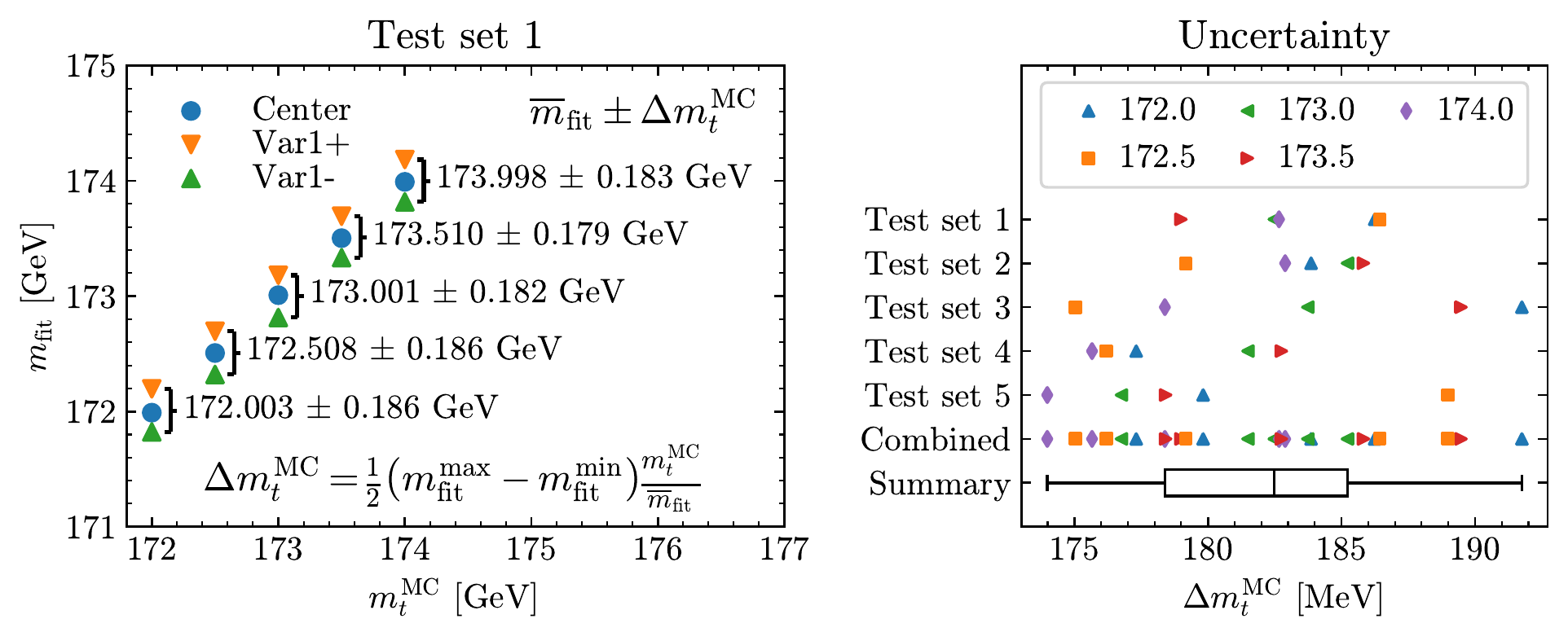}
    \caption{How uncertainties on $\mtmc$ are estimated. These particular numbers are from the uncorrected histogram fitting method using the Var1 tune, but the same error estimation is used throughout. Left:
    we show the fitted mass for different truth $\mtmc$ values and $+$ and $-$ variations of the tune parameters. For each $\mtmc$, the uncertainty $\Delta \mtmc$ is computed using equation~\ref{eqn:error estimate}.
    We repeat the fits with 5 test sets. Right: the values of $\Delta \mtmc$ for each test set and each mass are shown. The markers correspond to specific $\mtmc$ samples. The distribution of these uncertainties are shown in the box-and-whiskers plots (``summary" row of right panel).
    }
    \label{fig:DeltaExample}
\end{figure}

An example of this procedure is shown Fig.~\ref{fig:DeltaExample} for the Var1 tunes.
The blue, orange, and green data points denote the fits from the central, $+$, and $-$ variations, respectively.
The $x$-axis shows the true Monte Carlo mass of the sample and the $y$-axis gives the fit value.
We compute $\Delta \mtmc$ for five different values of the top mass: $\mtmc$=172.0, 172.5, 173.0, 173.5, and 174.0 \gev.
The spread between the maximum and minimum fit mass at each point is marked, and the average mass is reported along with $\Delta \mtmc$ for each Monte Carlo mass.
Note that the value of $\Delta \mtmc$ is different for each $\mtmc$.

In order to get a statistical estimation of the the uncertainty, we repeat the analysis on the same five masses using four more independent data sets generated with new random seeds.
The maximum and mean uncertainties from the 25 samples (5 masses times 5 data sets) are presented in the following figures. To visualize these uncertainties, we show box-and-whisker plots. These start by placing a box covering the 25th-75th percentiles of the $\Delta \mtmc$ values.
The whiskers then extend as a line out to the maximum and minimum, unless these are further away from the box than 1.5 box lengths, in which case the points are considered to be outliers and denoted by open circles. The line in the box denotes the median. An example of this statistical estimate is also shown in Fig.~\ref{fig:DeltaExample}.

\section{Classical fitting methods}
\label{sec:Classical}

In order to benchmark the tune uncertainties for the top mass, we first implement
two template fitting procedures modeled roughly on what is often done for actual experimental data.

\subsection{Histogram Fitting}
\begin{figure}[t]
    \centering
    \includegraphics[width=0.4\textwidth]{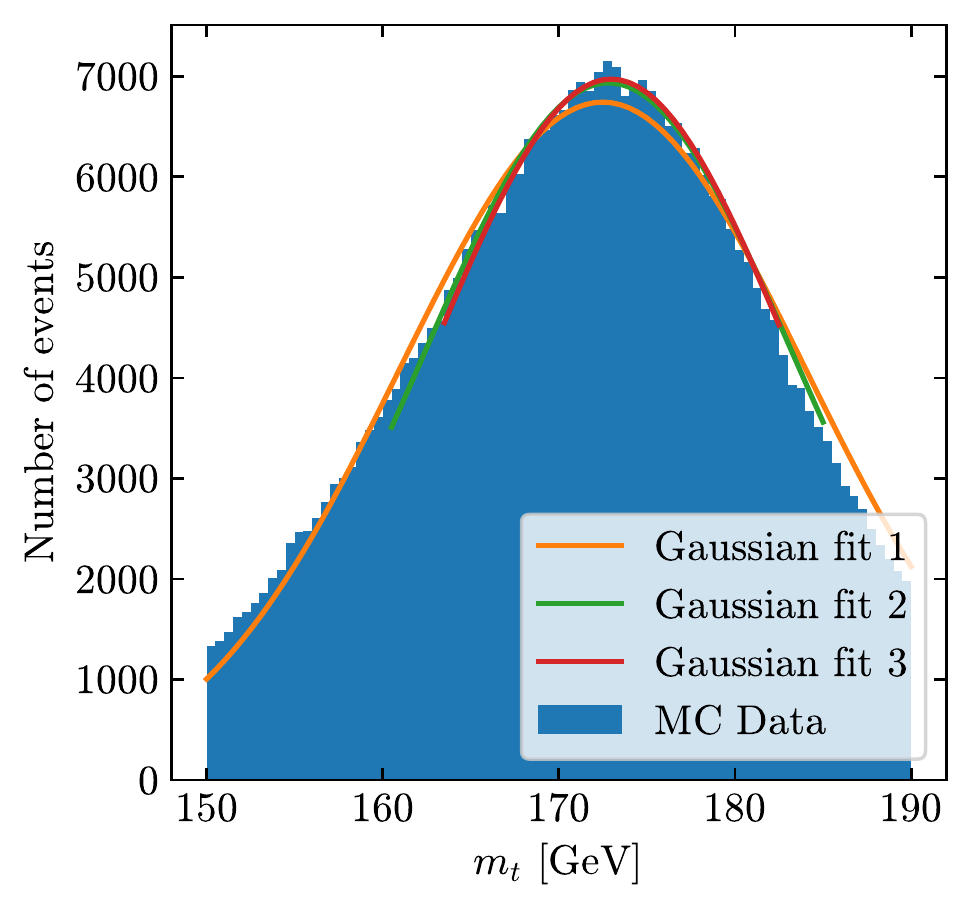}
    \includegraphics[width=0.4\textwidth]{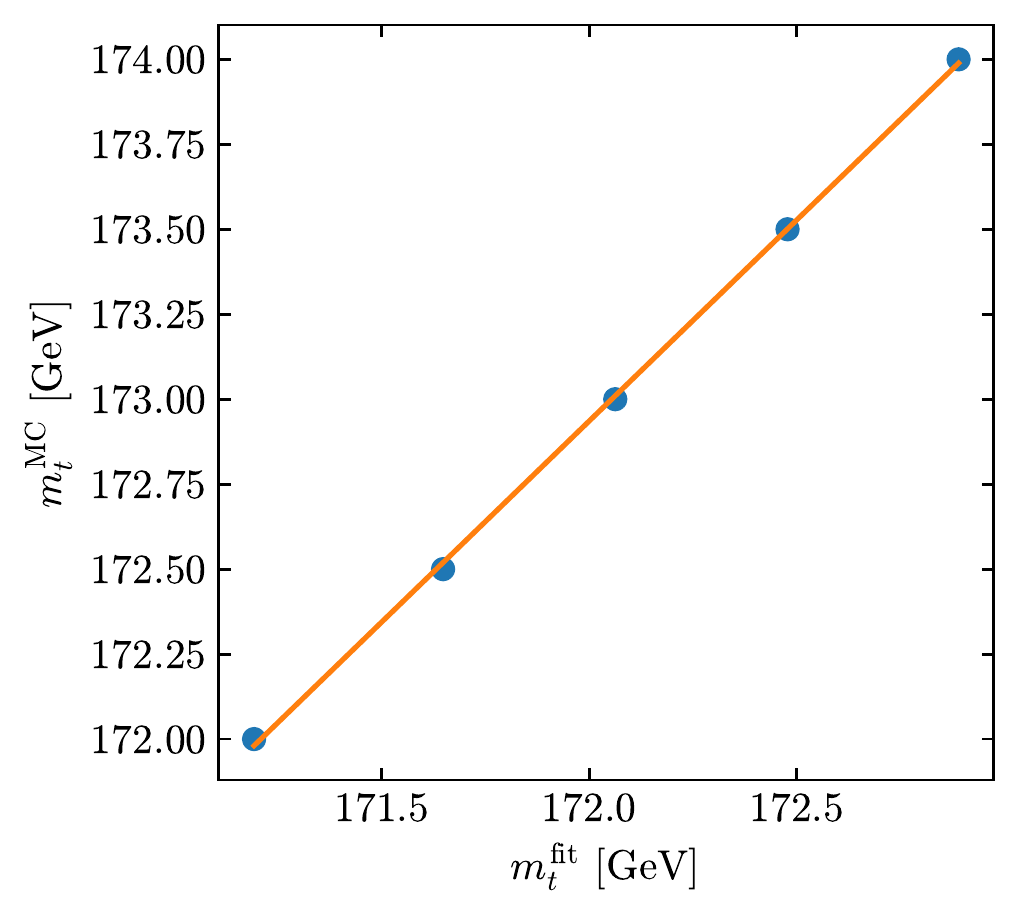}
    \caption{Shows the iterative fitting procedure used to fit a top mass to the 3-jet mass distribution. The left panel shows the distribution and several fits. In each iteration the fit range is adjusted to include one standard deviation on either side of the mean of the previous fit. The right panel demonstrates the linear relation between the fitted values of $m_t^{\text{fit}}$ from tune 21 data and the Monte Carlo mass $\mtmc$ used to generate the events. The
    fit ranges shown are $150-200$ GeV, then $159.4-185.6$ GeV and finally $162.3-183.1$ GeV.}
     \label{fig:iter_fit}
\end{figure}

We employ an iterated Gaussian histogram fit, similar to that used in~\cite{Andreassen:2017ugs}. For each test set at each tune and mass, we create a histogram of the three-jet invariant mass, $m_{3j}$, using anti-$k_T$ $R=0.5$ jets. We then fit a Gaussian to the distribution along the full range 150 GeV $< m_{3j} <$ 200 GeV. The fit range is then adjusted to include one standard deviation on either side of the mean of this Gaussian and a new Gaussian is fit to this new range. We continue to iterate this fitting procedure until the mean and width of the Gaussian converge to stable values. The mean is then used as the fitted top mass $m_t^{\text{fit}}$ and the width discarded. The left panel of Fig. \ref{fig:iter_fit} depicts this procedure. We also tested iterated fits of different functions such as a crystal ball function and a skewed gaussian, but do not display the results since they do not improve the top mass fit compared to the Gaussian case.  

The top mass $m_t^{\text{fit}}$ extracted from this method is very nearly linearly proportional to the top Monte Carlo mass $\mtmc$. This linear fit is shown on the right panel of Fig.~\ref{fig:iter_fit}. We then use the fitted mass and Monte Carlo mass to compute an uncertainty as described in the previous section. 

We consider three variants of this method, again following~\cite{Andreassen:2017ugs}. We first fit directly to the 3-jet mass histograms. Second, we calibrate to the $W$ mass. To do this, we rescale the 3-jet mass so that the 2-jet mass, $m_{2j}$, is equal to the $W$ mass:
\begin{equation}
m_{\text{calibrated}} = m_W\frac{m_{3 j}}{m_{2j}}.
\end{equation}
Finally, we apply the soft drop algorithm~\cite{Larkoski:2014wba} with parameters $\beta = 0, 1, 2$ on the jets before computing the histogram.

\begin{figure}[p]
    \centering
    \includegraphics[width=\linewidth]{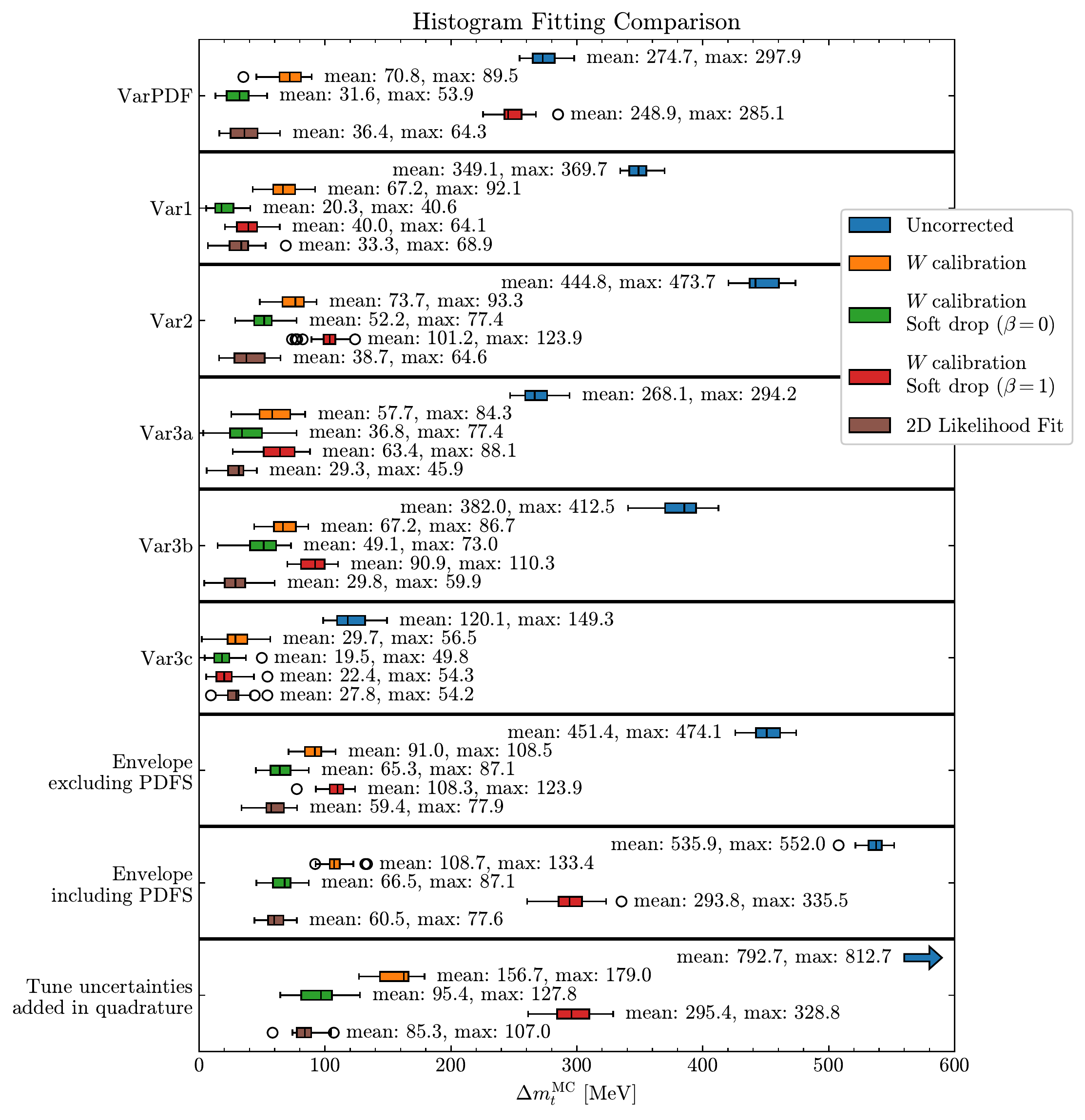}
    \caption{Box-and-whisker plot for the uncertainties on the top mass histogram fit and profile likelihood fit within each group of variations in the A14 set of tunes, as in Fig.~\ref{fig:DeltaExample}. The open circles denote outliers which are further away from the box than 1.5 box lengths. The uncertainties are calculated with and without $W$ calibration and soft drop grooming methods. The bottom three rows show several ways of combining the error for the different variations.}
    \label{fig:ClassicalMethods}
\end{figure}

A summary of the resulting uncertainties from the histogram-fitting approach is presented in Fig.~\ref{fig:ClassicalMethods}. We find that the best variant, including both $W$ calibration and soft drop with $\beta = 0$, yields a mean total envelope uncertainty of about 65 MeV and an uncertainty of about 100 MeV when the variations are added in quadrature.  This is roughly consistent with the values in~\cite{Andreassen:2017ugs}, and similar to values found by ATLAS~\cite{Aaboud:2018zbu}.

\subsection{Profile likelihood fitting \label{sec:2dlikelihood}}
The histogram fitting method does not easily extend to more than one observable and does not include information from every event.
A second method which is used by the experiments is to perform a profile likelihood fit.
The idea behind this method is to find the mass which is most likely to have generated the observed events.
To do so, a likelihood function is used to model the distribution.
The likelihood function is able to incorporate more than one observable, allowing for more flexibility than the histogram fitting.

Here, we model the top and $W$ resonances as Gaussian distributions.
The mean value of the top distribution will be fit, and the mean of the $W$ distribution is set to $80.3\gev$. 
The standard deviations of the distributions are determined from fitting the resonances across all tunes simultaneously and are $\sigma_t=6.5\gev$ and $\sigma_W=3.5 \gev$.
In addition, we include a nuisance parameter, $c$, in the model to help account for fluctuations in the ratio of the reconstructed top and $W$ masses coming from differences in the tune parameters.
Explicitly, the likelihood is given as
\begin{equation}
    \mathcal{L}\left(m_t^{\rm{fit}}, c\right) = \prod_{i\in \rm{Events}} \bigg(\mathcal{G}\left(m_{3j} \, c ~| m_t^{\rm{fit}}, \sigma_t \right) 
    \mathcal{G}\left(m_{2j} \, c ~| m_W, \sigma_W \right)
    \mathcal{G}\left(c ~|1, \sigma_c \right) 
    \bigg)~,
    \label{eqn:Likelihood}
\end{equation}
where $\mathcal{G}(x|\mu, \sigma)$ is the probability density evaluated at $x$ for Gaussian distribution with mean $\mu$ and standard deviation $\sigma$, and $\sigma_c$ is the standard deviation of the fitted value of the $m_{2j}$ peaks across a range of tunes. 
We use $\sigma_c = 0.13$.

For a set of events with a fixed tune, the value of the top mass is extracted by maximizing the likelihood function over both $m_t^{\rm{fit}}$ and $c$.
The value of $m_t^{\rm{fit}}$ that maximizes the likelihood does not equal $\mtmc$, but is linearly correlated.
The linear relation between $\mtmc$ and $m_t^{\rm{fit}}$ is used for the inference of top mass.

Fig.~\ref{fig:ClassicalMethods} shows the results using this method as the brown bars denoted by ``2D Likelihood Fit".
Overall, this method does similar to the histogram fitting with grooming and calibration, even though these are not done explicitly here.
The likelihood fit improves the mean $\Delta\mtmc$ by around 10\% when taking the envelope of the tunes or adding the uncertainties in quadrature.
This improvement comes as a result of using the values of $m_{2j}$ and $m_{3j}$ from every event and including a nuisance parameter.
In principle, it is possible to implement a nuisance parameter for each of the tuning parameters, but this is challenging in practice, as the effects of each tuning parameter may not be well modeled by a Gaussian.
Instead, we advocate for the method presented in the next section, which still includes the values of $m_{2j}$ and $m_{3j}$ from every event, but allows for a flexible function--unlike the fixed form of Eq.~\eqref{eqn:Likelihood}--which uses changes in the distributions to account for tune variations. 

\section{Regression on Sorted Ensembles \label{sec:dense}}

In this section, we study whether doing regression on 
ensembles of events can improve on the traditional template histogram fit.
We consider both using a dense neural network (DNN) to do the regression with machine learning, and alternatively an ordinary least squares (OLS) linear regression. 
The two methods give comparable results. The DNN with linear activations is slightly better, but the OLS regression is simpler and faster (but uses more memory). 

\subsection{Inputs}
We take as inputs to the regression ensembles of observables computed from simulated events, sorted by one of the observables. We use these sorted ensembles as inputs to regress out the top-quark Monte Carlo mass, $\mtmc$. 
 
For training, we use events simulated with  $\mtmc$ ranging from 170 to 176 GeV in intervals of 0.2 GeV. For each mass we generate 300,000 training events for each of the A14 tunes. We have also tested using a larger total number of events in each sample and finer spacings between the masses, but this does not improve our results. There is no apriori reason why a uniform prior necessarily gives the best performance, but we find it to be sufficient for our purpose. To train the regression, we use a random ensemble of 30,000 events (with replacement) from the total set of 300,000 at a given tune and $\mtmc$. The number 30,000 is chosen because taking a smaller number of events per ensemble gives a larger error, while taking a larger number is prohibitively slow (at least in the DNN case) and does not lead to noticeable improvement. Both the DNN and OLS regression see many different ensembles from each training sample, but the total number of ensembles and which samples they are from differs between the two regression methods. For the DNN, batches of 100 ensembles are seen in each training step, and each ensemble is from a randomly selected mass and tune. In contrast, the best OLS regression uses 20 ensembles for each mass and tune.

\begin{figure}[t]
     \centering
     \includegraphics[width=\textwidth]{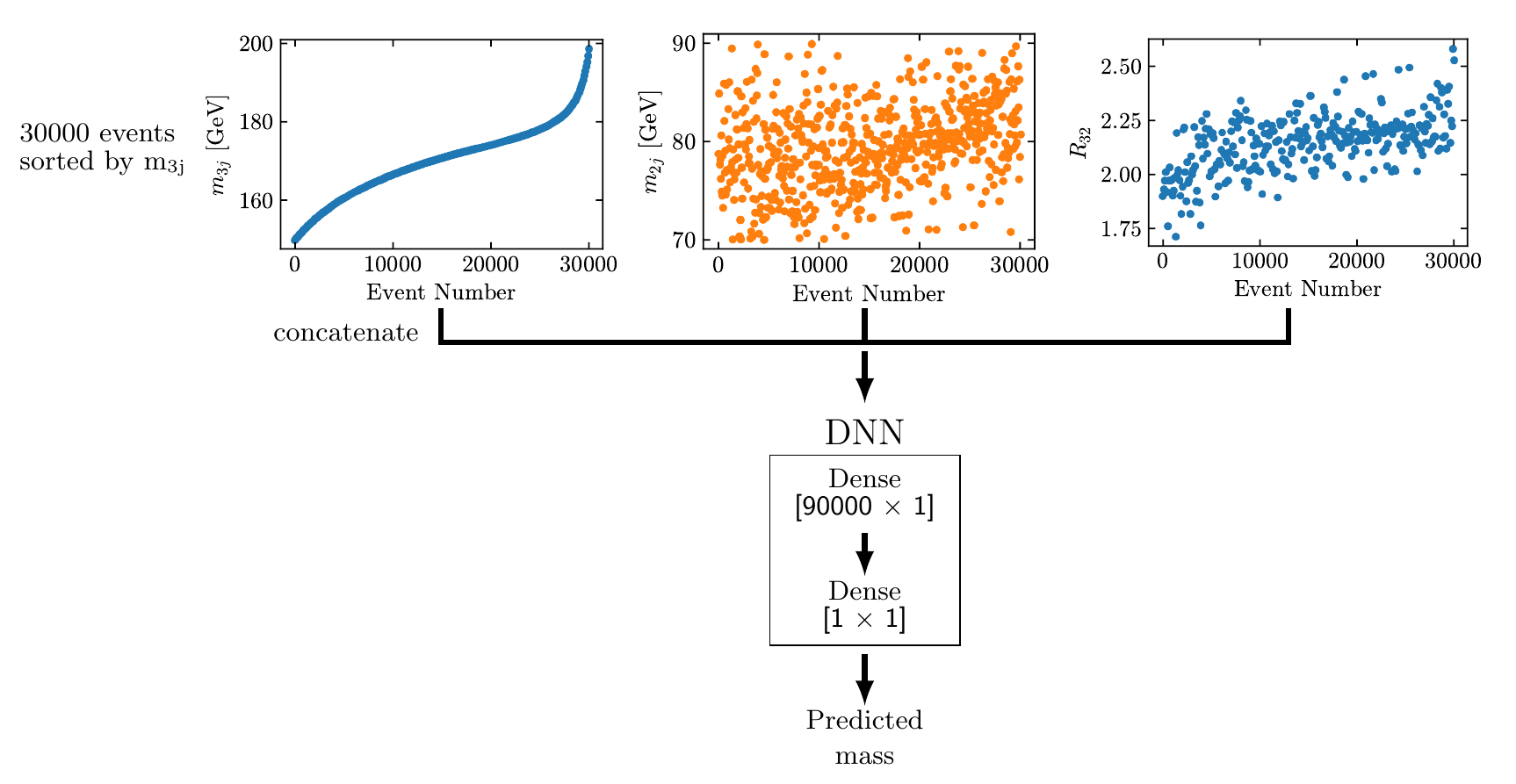}
     \caption{Example DNN and its inputs. Example inputs are $m_{3j}$, $m_{2j}$, and $R_{32}$ for 30,000 events, sorted according to increasing $m_{3j}$.}
     \label{fig:DNN_sort}
\end{figure}

The basic observables we consider are the 3-jet invariant mass (i.e. top), $m_{3j}$, the 2-jet invariant mass (i.e. the $W$ boson mass), $m_{2j}$, and their ratio $R_{32}=\frac{m_{3j}}{m_{2j}}$~\cite{Argyropoulos:2014zoa,Skands:2007zg}. The inputs to the regression are the values of these observables, sorted by one (or more) of them. 
Sorting the ensemble is important because it determines which parts of each observables' distribution the different weights are applied to, and allows the regression to exploit correlations in different observables across tunes.
Example input distributions sorted by $m_{3j}$ are shown in Fig.~\ref{fig:DNN_sort}.
We tested different orderings and several different observables as inputs (discussed more in section \ref{Section: DNN Results}), and find that sorting by increasing $m_{3j}$ tends to give the best results.

To extract the uncertainty from the regression, we generate five more statistically independent samples of 400,000 events at each mass between 172.0 and 174.0 GeV (in intervals of 0.5 GeV) and for each of the A14 tunes. From each of these test samples (at fixed mass and tune), we take an ensemble of 30,000 events and evaluate the network to get an output value. We repeat this 100 times for each sample and take the mean of those values to get a final predicted value for a given trial, mass, and tune. We then use those predicted values to compute the error as described in section \ref{sec:evtgeneration}.

\begin{figure}[p]
    \centering
    \includegraphics[width=0.85\textwidth]{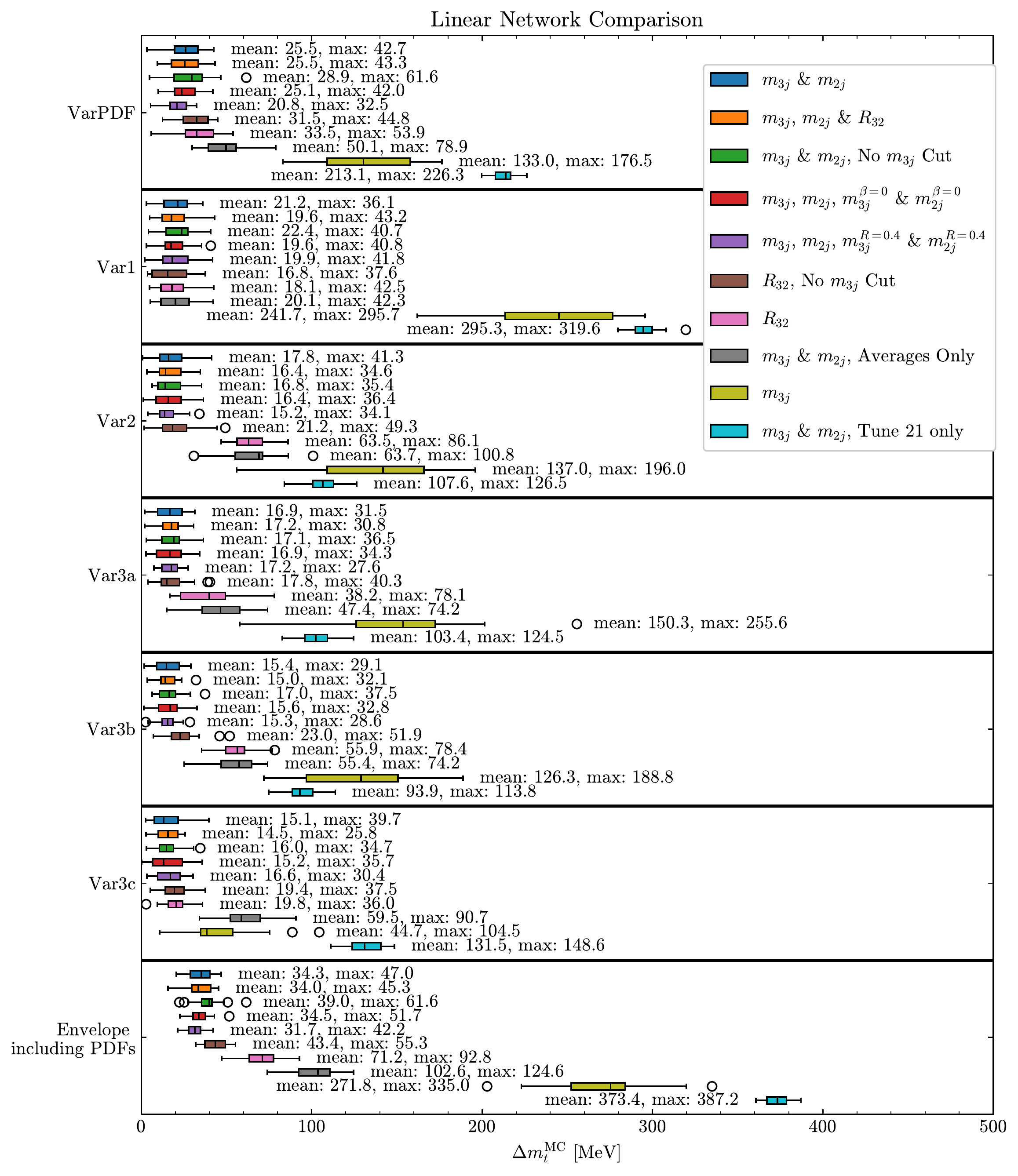}
    \caption{{\small{
    Uncertainties on the top mass linear network fits within each group of variations in the A14 set of tunes, with different observables as inputs.
    The Total Envelope section contains the envelope of all tunes. 
    For every network displayed (except that trained only on the means of the distributions and the soft drop example), the distributions are sorted by increasing $m_{3j}$.
    For the soft drop example, the distributions are sorted by $m_{3j}$ with $\beta = 0$ rather than the original $m_{3j}$. We have restricted $m_{3j}$ to between 150-200, except for those networks labeled ``No $m_{3j}$ cut".}}}
    \label{fig:DNN_Compare_Final}
\end{figure}

\subsection{Dense Network
\label{Section: DNN Results}}

First, we discuss using a linear network.  We use a two layer network implemented in {\sc keras}~\cite{chollet2015keras}, with one node in each layer and linear activation functions between nodes.\footnote{The exceptions to this are the $R_{32}$ only networks, which train better when we use a third layer.} This is shown in Fig.~\ref{fig:DNN_sort}. While more than one layer is not strictly necessary since our activation functions are linear, additional layers can help with training and hyperparameter optimization.
We also tested more complicated neural networks with different filter configurations (including deeper networks), removing various node connections, and nonlinear activation functions, none of which seemed to improve performance. Networks were trained with the Adam algorithm~\cite{kingma2017adam} for 600 epochs of 750 steps each, with an early stopping patience of 60 and a batch size of 100. The initial learning rate was 0.0005, with a learning rate decay of 0.7 after 8 epochs without improvement. We did not exhaustively optimize these hyperparameters, so it is possible that there would be further performance gains with a more systematic hyperparameter search.  We also normalize all inputs by subtracting a constant so that the mean of each sorted ensemble is small compared to its spread, which helps ensure consistent results when the network is trained multiple times. This amounts to subtracting 173 GeV from $m_{3j}$, 80 GeV from $m_{2j}$, and 2 from $R_{32}$. We also tested other normalization methods, but found they do not improve performance noticeably.
We tested several loss functions and determined that the network is mostly insensitive to which loss function was used and performs equivalently for loss functions such as logcosh and mean squared error. The results presented use the logcosh loss. 

We tested multiple different sets of observables as inputs. We considered including combinations of $m_{3j}$, $m_{2j}$, $R_{32}$, and $m_{\ell b}$ (the invariant mass of the lepton and b-quark on the leptonically decaying top quark side of the event). For each of these inputs, we considered sorting the ensembles in different ways before putting them into the network. Sometimes, we sorted the events the same way for all the observables, and sometimes we sorted the events differently for different observables. In either case, the orderings were determined by making sure the sorted variables were strictly increasing, and then applying one of these orderings to each of the other, unsorted variables. We find sorting is necessary to train the network, and that the results depend on how the inputs were sorted. Generically, we find that sorting by $m_{3j}$ works best.

\begin{figure}[p]
    \centering
    \begin{subfigure}[b]{0.48\textwidth}
         \centering
         \includegraphics[width=\textwidth]{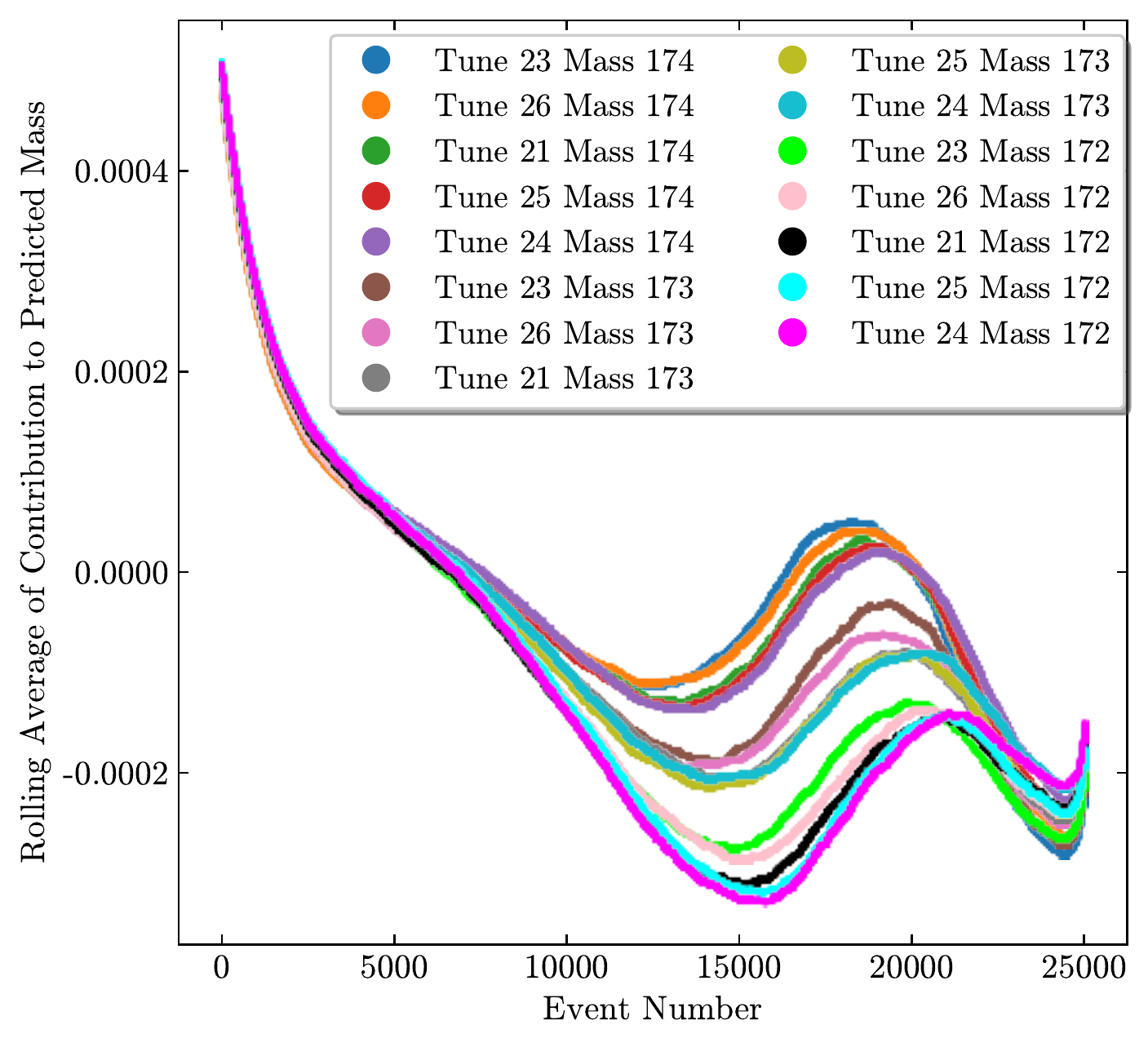}
        \captionsetup{justification=raggedleft, singlelinecheck=false}%
        \caption{\phantom{SSSSSSSSsssspace}}
         \label{fig:M2J_M3J_incremental}
    \end{subfigure}
    \hfill
    \begin{subfigure}[b]{0.45\textwidth}
         \centering
         \includegraphics[width=\textwidth]{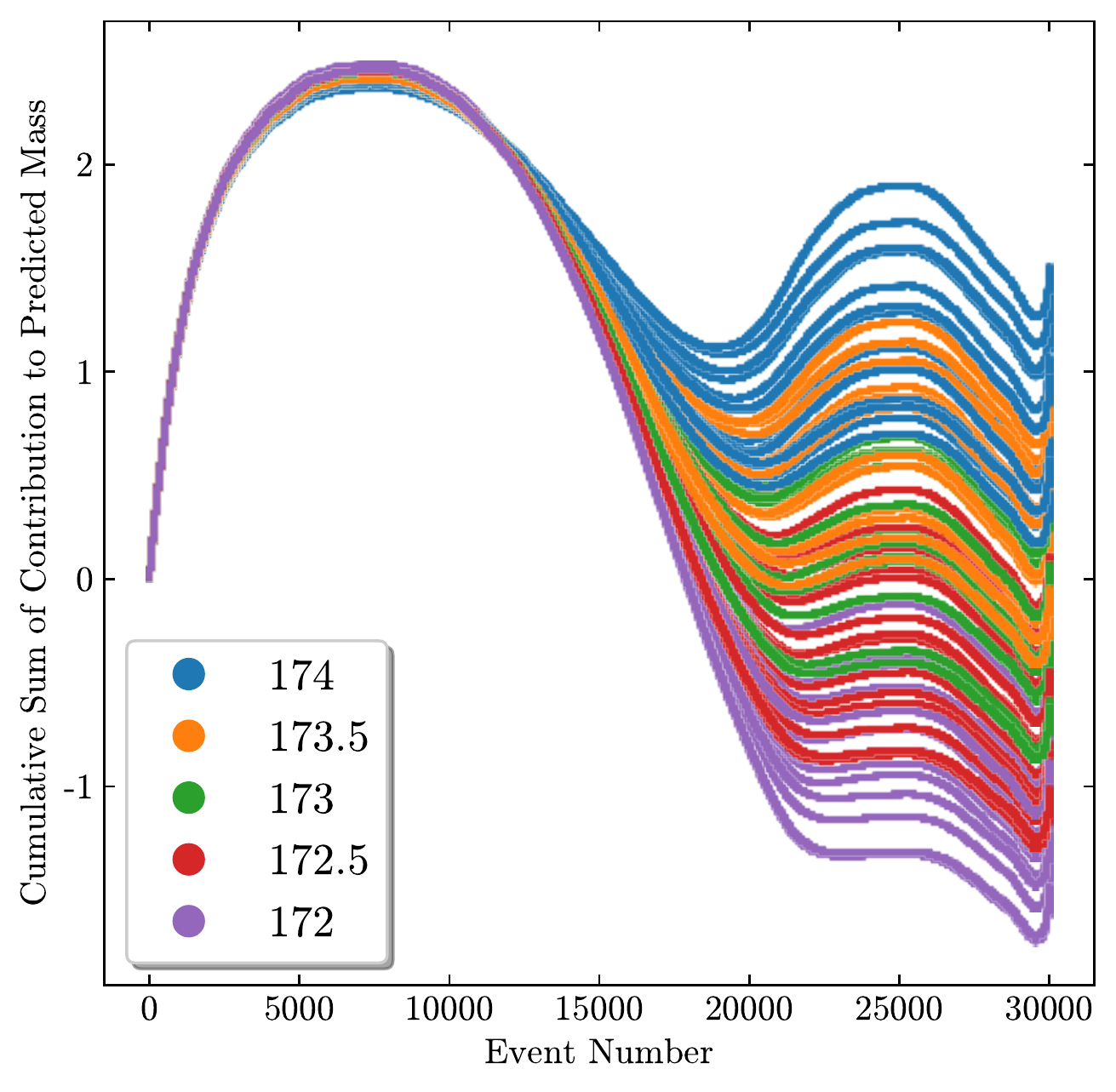}
        \captionsetup{justification=raggedleft, singlelinecheck=false}%
        \caption{\phantom{SSSSSSSSSSspace}}
        
        \label{fig:M2J_M3J_cumulative_M3J_only}
     \end{subfigure}
      \vfill
     \hskip 0pt plus 0.3fill 
     \begin{subfigure}[b]{0.45\textwidth}
         \centering
         \includegraphics[width=\textwidth]{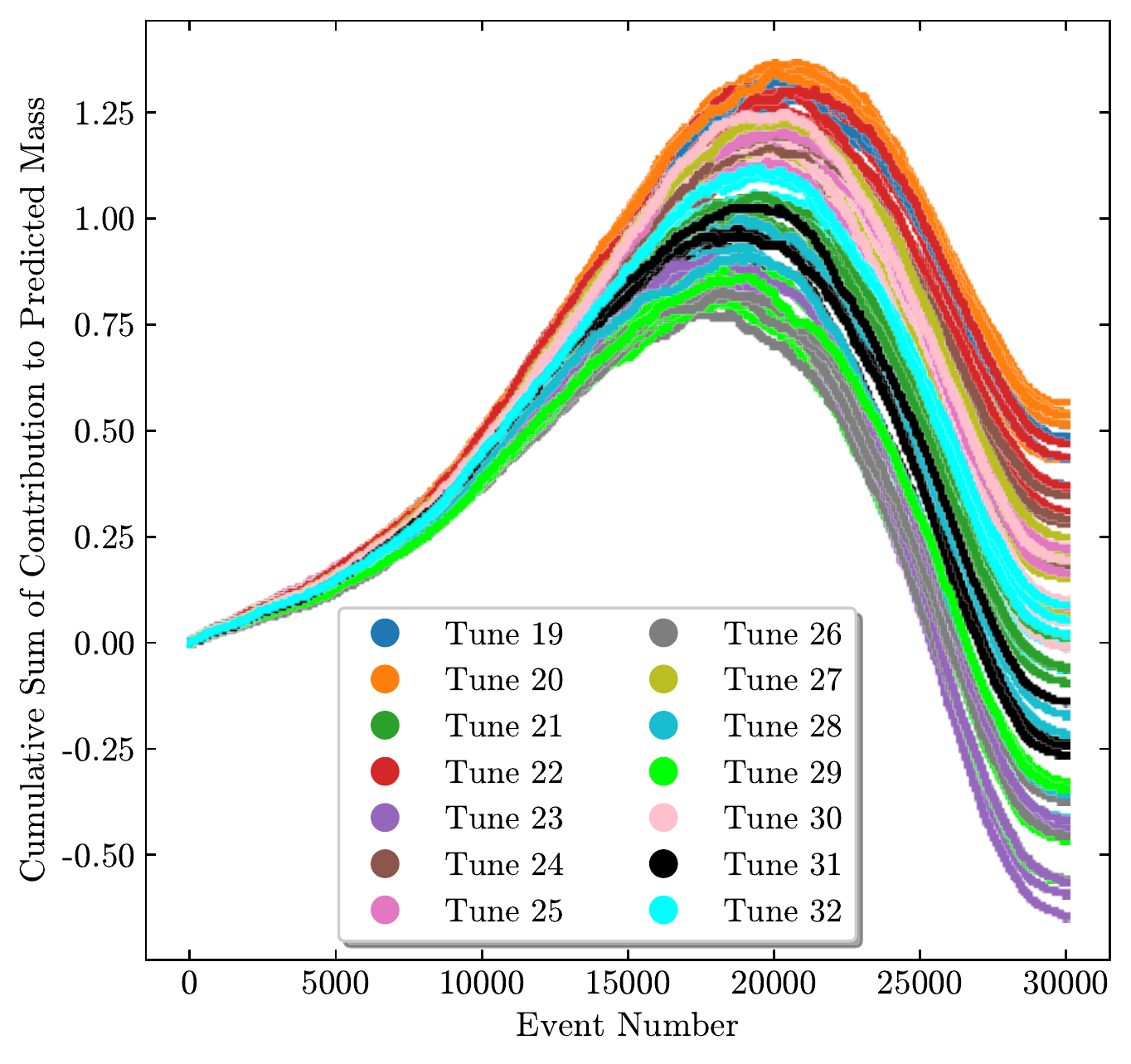}
        \captionsetup{justification=raggedleft, singlelinecheck=false}%
        \caption{\phantom{SSSSSSSSSSSpce}}
        
        \label{fig:M2J_M3J_cumulative_M2J_only}
     \end{subfigure}
    \hskip 0pt plus 0.7fill
    \begin{subfigure}[b]{0.45\textwidth}
         \centering
         \includegraphics[width=\textwidth]{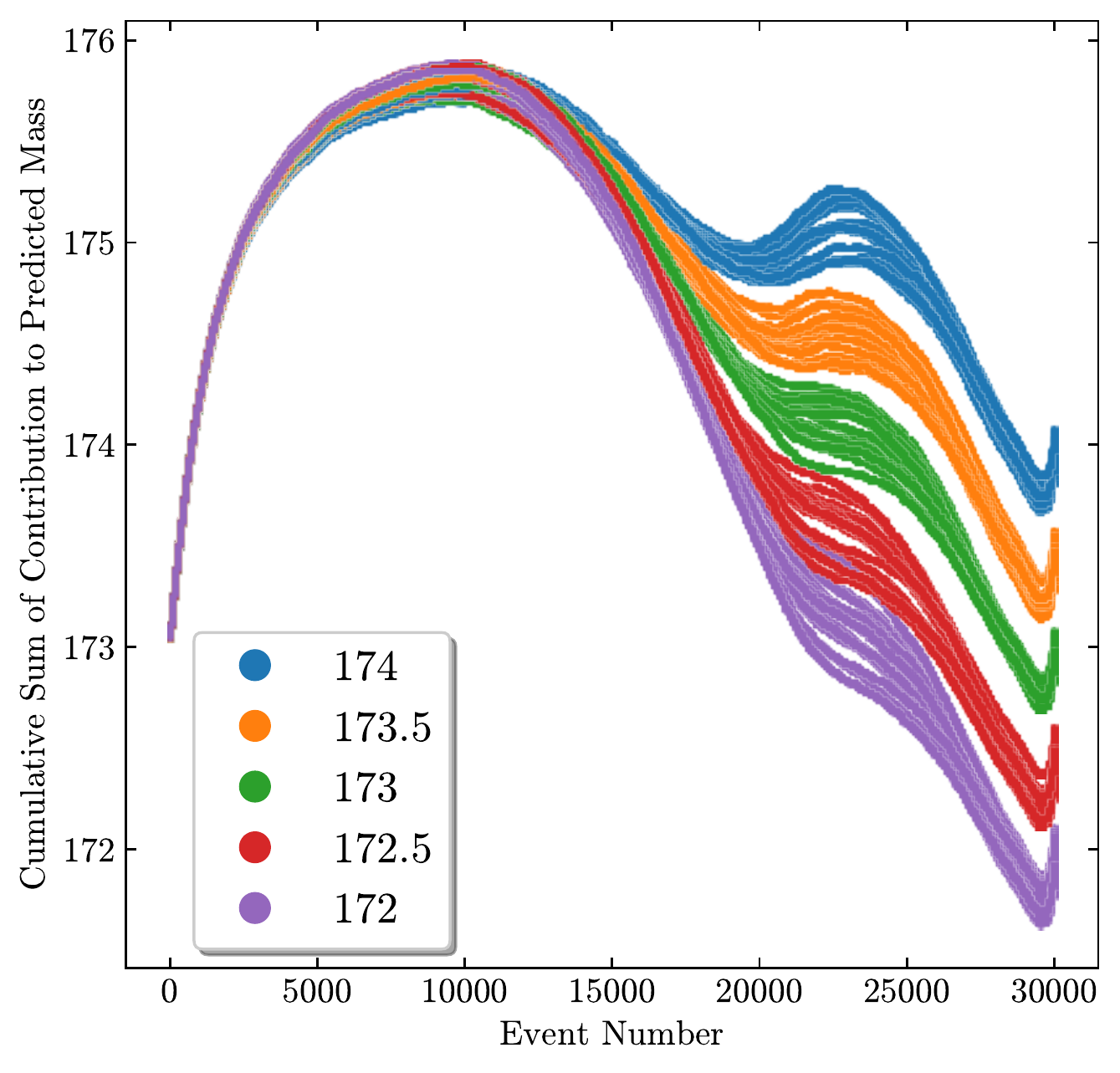}
        \captionsetup{justification=raggedleft, singlelinecheck=false}%
        \caption{\phantom{SSSSSSSSSSSpce}}
        
    \label{fig:M2J_M3J_cumulative_both}
     \end{subfigure}

        \caption{Graphical representations of the contribution to the predicted mass as a function of entry number in the ensemble using the $m_{3j}$ \& $m_{2j}$ network. In (a), we show a rolling average of the product of the (normalized) $m_{3j}$ input multiplied by the network weight for each event number. Each color denotes a different mass and tune. Only a subset of masses and tunes have been shown for clarity. In (b)-(d), each point is the cumulative sum up to that event number of the (normalized) input value multiplied by the network weight. (b) includes only the $m_{3j}$ contribution, with colors denoting masses and each mass including 1 sample from each of the A14 tunes. (c) includes only the $m_{2j}$ contribution, with colors denoting tunes. One sample at each mass is included for each tune. (d) sums over both $m_{3j}$ and $m_{2j}$ with colors again denoting masses. The bias is included in the zeroth entry.}
        \label{fig:M2J_M3J_Weights}
\end{figure}

Additionally, we recomputed these observables for subjets determined in different ways and tested combining these subjet observable ensembles with those for the original jets. We used subjets obtained by applying soft drop \cite{Larkoski:2014wba} with $z = 0.1$ and $\beta = 0,1,2$ to the initial jet, as well as telescoping subjets at different radii~\cite{Ellis:2012sn,Chien:2013kca,Chien:2014hla}. We found that our results are not sensitive to the value of $z$ used in soft drop as long as it is small enough. Changing $\beta$ has a small effect which is not noticeable in the best case of $m_{3j}$ \& $m_{2j}$ inputs. For single observable networks, $\beta = 0$ often does best. For both types of subjets, our results depend on the specific network inputs, but none of these networks perform noticeably better than the best network without subjet observables.

A subset of our results for various different inputs is shown in Fig.~\ref{fig:DNN_Compare_Final}. As previously mentioned, a combination of $m_{3j}$ and $m_{2j}$, both sorted by increasing $m_{3j}$, is sufficient to give our best results. This is shown in blue. From the figure, we can see that most of the contributions from the different variations are a similar order of magnitude, in contrast to the histogram fitting case. The largest contributions to the error are from Var1 and VarPDF.  

For completeness, Fig.~\ref{fig:DNN_Compare_Final} contains several other results, considering both different input variables and different $m_{3j}$ ranges. We find many networks with additional variables perform similarly to the $m_{3j}$ \& $m_{2j}$ combination (which can be seen by comparing the blue errors to the orange and green ones), while networks that do not include both $m_{3j}$ and $m_{2j}$ tend to perform worse. Like in the case of the histogram fit, using mass alone (shown in yellow) gives the worst results, while $R_{32}$ alone (shown in pink and brown) improves upon the mass, though both do better than the histogram fit when the same inputs are used.\footnote{The improvement between $R_{32}$ and using multiple observables is dependent on the range of $m_{3j}$ used; for $m_{3j}$ between 150-200 the difference is larger than when the full $m_{3j}$ distribution is used. This can be seen in the difference between the pink and brown errors.} Examples including soft drop variables and subjets at different radaii are shown in red and purple respectively. Including soft drop variables or subjets at smaller radii can help when compared to networks trained on single variables, but there is no further improvement on the $m_{3j}$ \& $m_{2j}$ combination. For reference, we also include two other networks that do not use the full distributions for every tune as input. In gray, we show the case of taking the average of the ensemble before inputting to the network. In turquoise, we show a network trained on tune 21 only (but still tested on all the A14 tunes). Unsurprisingly, we find that in both of these cases the networks perform worse than the full sorted ensembles marginalized across tunes.

Next, we would like to understand why these networks are able to perform better than the histogram fit. In the histogram fit, the mass is given by the center of a Gaussian which is similar to our average-only network. Therefore, we look for improvements over the average-only network as a proxy for understanding why the dense network does better than the histogram fit. In order to understand how different parts of the ensembles contribute, we examined the weights of the networks. 

\begin{figure}[t]
     \centering
     \begin{subfigure}[b]{0.45\textwidth}
         \centering
         \includegraphics[width=\textwidth]{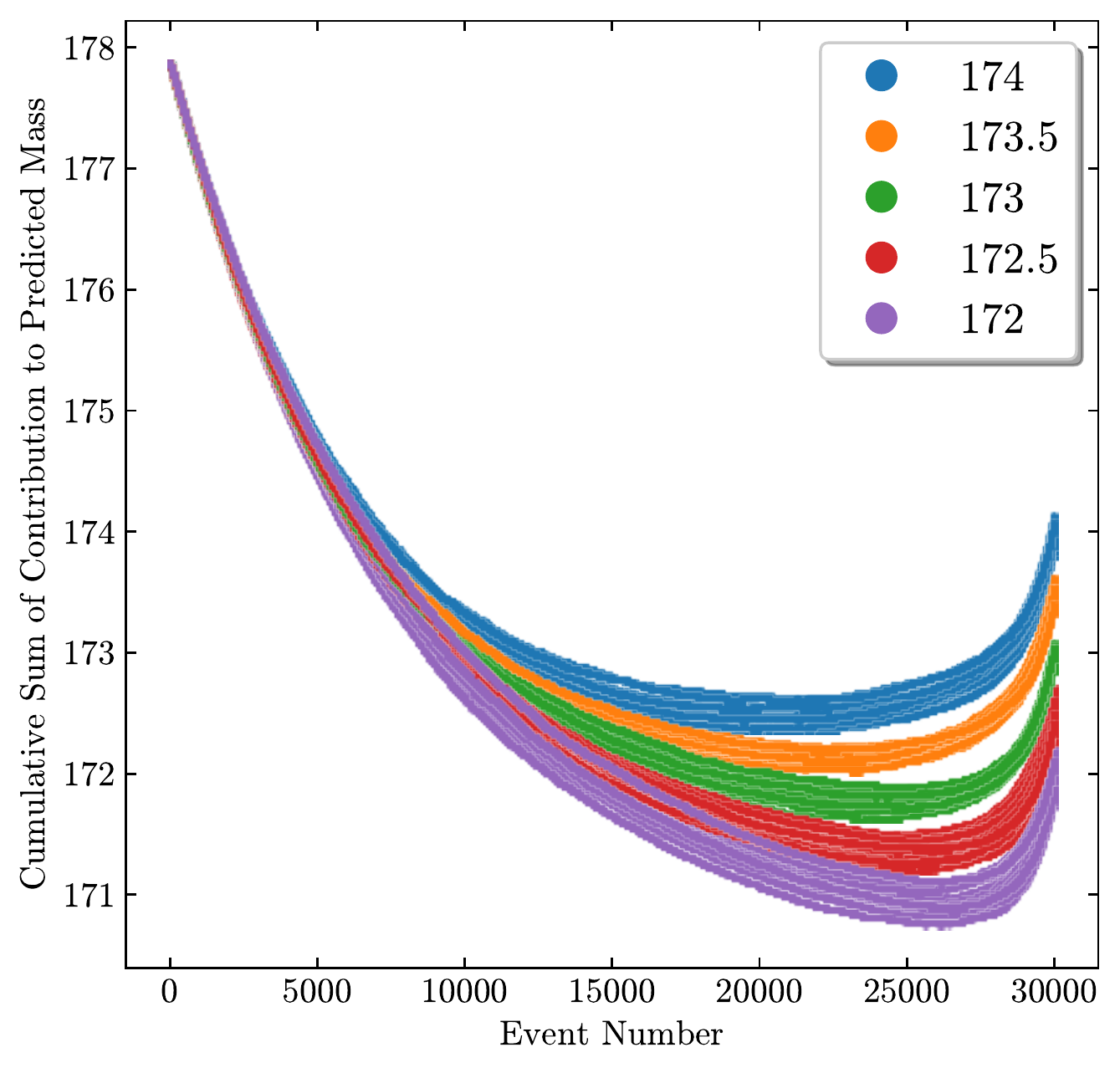}
        \captionsetup{justification=raggedleft, singlelinecheck=false}%
        \caption{\phantom{SSSSSSSSSsspace}}       \label{fig:M3J_M2J_Means_Weights}
     \end{subfigure}
     \hfill
     \begin{subfigure}[b]{0.45\textwidth}
         \centering         \includegraphics[width=\textwidth]{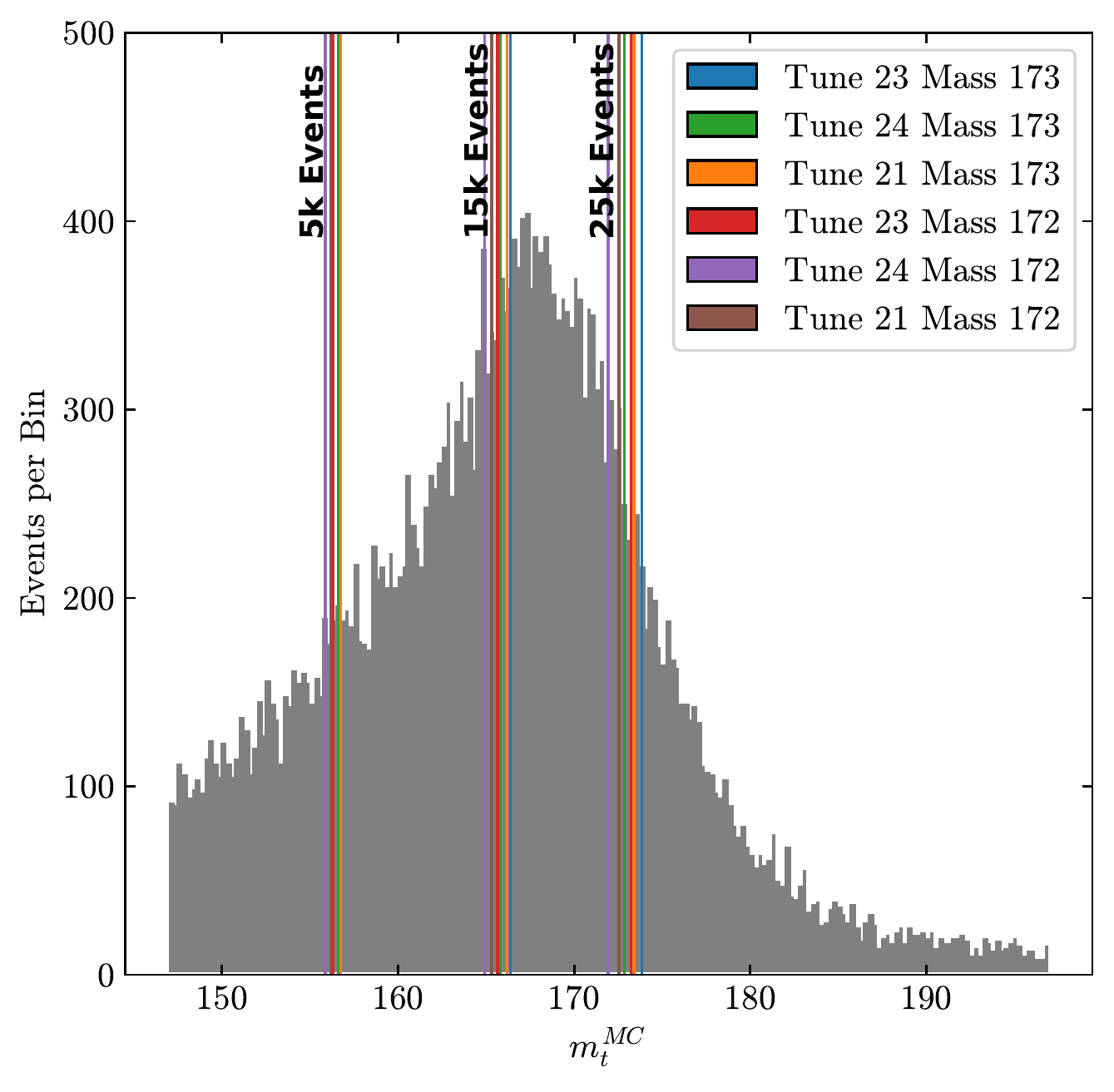}
        \captionsetup{justification=raggedleft, singlelinecheck=false}%
        \caption{\phantom{SSSSSSSSSsspace}}
         \label{fig:M3J_HIST}
     \end{subfigure}
        \caption{(a): Graphical representation of the contribution to the predicted mass as a function of entry number in the sorted ensemble using the network that depends only on the average of the ensemble. The zeroth entry includes the bias. Each point is the cumulative sum up to that entry of the (normalized) input values multiplied by the network weight and divided by the total number of samples. Color denotes the mass. There are fourteen different lines for each color; one example for each tune. (b): Histogram showing an example distribution. The solid lines drawn show 5k, 15k, and 25k events for several masses and VAR1 tunes.}
        \label{fig:means_only_+_hist}
\end{figure}

We use these weights to construct Figs.~\ref{fig:M2J_M3J_Weights} and \ref{fig:M3J_M2J_Means_Weights}. Since these weights are shared by all masses and tunes, we multiply the weights by example input ensembles to construct the plots. Specifically, we plot a rolling average of the input times the weights (as in Fig.~\ref{fig:M2J_M3J_incremental}) or a cumulative sum of the inputs times the weights (as in Figs.~\ref{fig:M2J_M3J_cumulative_M3J_only}-\ref{fig:M2J_M3J_cumulative_both} and  \ref{fig:M3J_M2J_Means_Weights}) as a function of event number. For Figs.~\ref{fig:M2J_M3J_cumulative_both} and \ref{fig:M3J_M2J_Means_Weights}, which include all input observables, we have also added the constant bias learned by the network and accounted for the normalization of the labels by adding 173 to the predicted outputs. In these two plots, we can read off the predicted $\mtmc$ value from event number 30,000. We include Fig.~\ref{fig:M3J_HIST} for reference, to see which parts of the top mass distribution are contributing the most in each network.

In general, we want to design a procedure that is sensitive to the Monte Carlo mass but not the tuning parameters. The difficulty with this is that most variables that are strongly affected by $\mtmc$ (such as $m_{3j}$) are also strongly affected by the tuning parameters, which we can see from Fig.~\ref{fig:M2J_M3J_cumulative_M3J_only}. This can be partially corrected by including other variables (such as $m_{2j}$, seen in figure~\ref{fig:M2J_M3J_cumulative_M2J_only}) which are more sensitive to the tune than the MC mass. When we just fit the mean of each distribution, there is not much more that we can do, aside from trying to clean up the distributions themselves. However, in the case of directly inputting a sorted ensemble into the regression, the network can look for other combinations that are less sensitive to the tune than the mean. The network can learn to use a particular part of an observables' distribution to differentiate the masses, and a different part to partially correct for the difference in tunes. This can be seen in Fig.~\ref{fig:M2J_M3J_incremental}, where the middle of the ensemble distinguishes the mass, whereas the upper tail is more strongly correlated with the tune. We can also see this from comparing the average only network in Fig.~\ref{fig:M3J_M2J_Means_Weights} with the full $m_{3j} \& m_{2j}$  network in Fig.~\ref{fig:M2J_M3J_cumulative_both}. In~\ref{fig:M3J_M2J_Means_Weights}, most of the difference in masses comes from events between 5,000-15,000, where the original distributions differ most, and the width of each mass band in the upper half of the ensemble remains mostly constant. In contrast, in~\ref{fig:M2J_M3J_cumulative_both}, most of the difference in regressed mass comes from event numbers greater than 15,000, and the difference in tunes shrinks substantially at the top tail of the $m_{3j}$ ensemble.

For completeness, we also tried generating new samples uniformly spaced in the other tuning parameters and regressing out these tuning parameters in addition to the mass. For this test we restricted to the VAR1 tunes, but an equivalent test could be conducted across all variations. We might think this type of network would improve our results since the loss function explicitly depends on tuning parameters in addition to $\mtmc$.  However, we found that in the case of a linear regression, a multidimensional output did not help improve the predicted top mass (in contrast to what we found with the DCTR method, discussed in Section~\ref{sec:DCTR}). In particular, we find that sorting the inputs encodes enough information about the other tuning parameters that additional outputs are unnecessary. This can be seen from the solid lines in Fig.~\ref{fig:M3J_HIST}. While the value of the 15,000th event near the peak depends primarily on the Monte Carlo mass, the value of the 25,000th event is also strongly dependent on the tune. 

\begin{figure}[!t]
    \centering
    \includegraphics[width=0.6\textwidth]{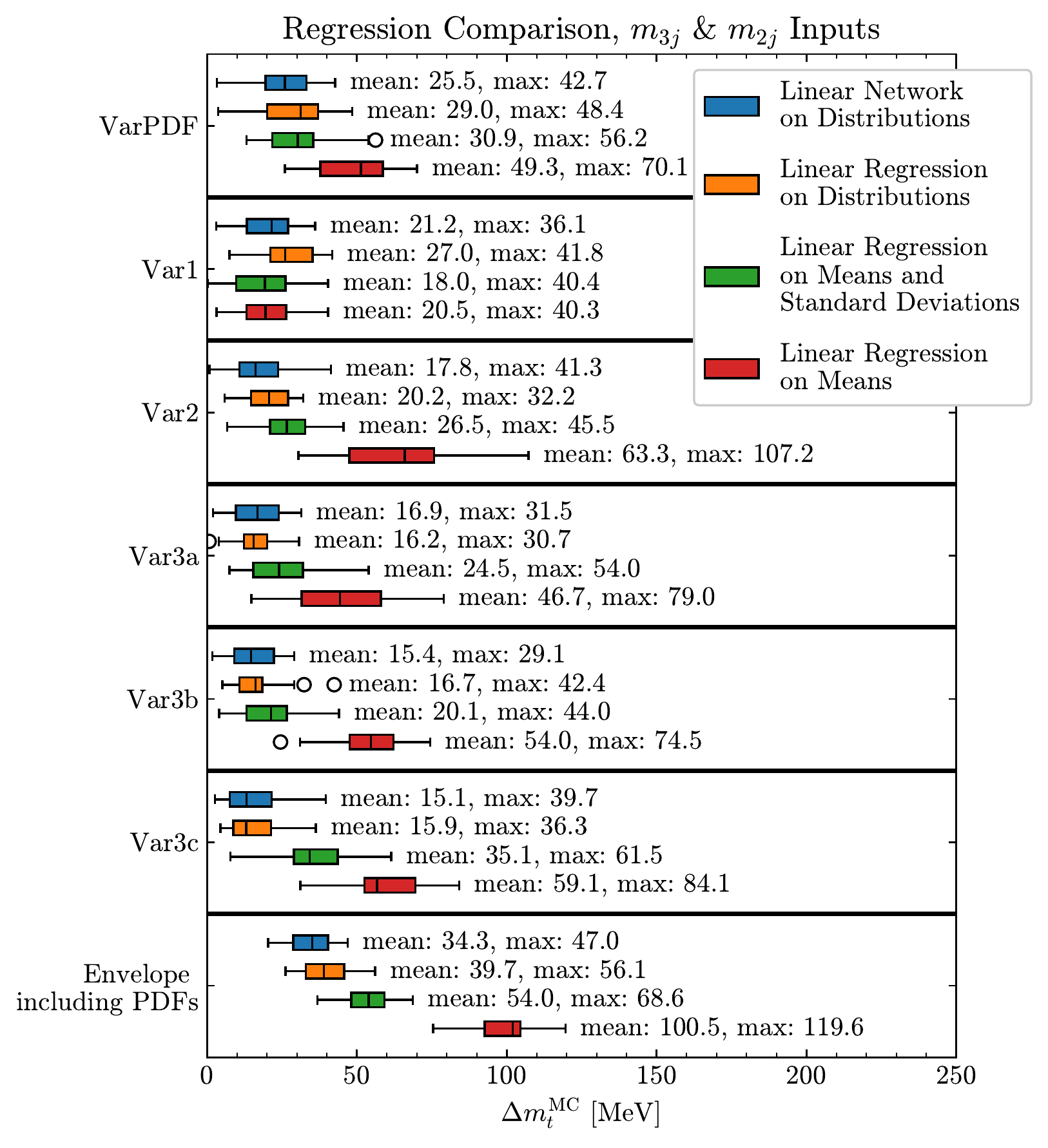}
    \caption{Uncertainties on the top mass fit within each group of variations in the A14 set of tunes for the OLS regressions, compared to the best linear network.}
    \label{fig:Regression_Compare}
\end{figure}

\subsection{Ordinary Least Squares Regression}

Since non-linearities and a deep network structure do not seem to improve results, it is natural to ask if we can reproduce the same results with something simpler. Therefore, we test the case of using a projection matrix to do the ordinary least squares linear regression exactly (rather than using the Adam algorithm to do the minimization). We implement this regression using scikit-learn\cite{scikit-learn}, and consider three separate cases: using the full ensembles of 30,000 events, using the means across the ensemble only, and using the ensemble means and their standard deviations. For the full ensembles we use 20 random samples from the 300,000 event training sets with each mass and tune; for the other cases we resample 200 times from each mass and tune. We find that with full ensembles the results are worse when fewer samples are used, but including more than 20 samples decreases performance. We suspect this is due to overparameterization, but that it could be improved through regularization or dimensionality reduction techniques. Additionally, the matrix operations in the OLS regression become memory intensive  (using over 32 GB) with many samples. The other regression methods are mostly insensitive to the number of samples, as long as there enough for fitting. Results from the OLS approach are shown in Fig.~\ref{fig:Regression_Compare}. We find roughly similar, but slightly inferior, performance to the linear network for comparable inputs. Though the linear network does slightly better, it takes longer to train than OLS regression and is not deterministic.

\section{DCTR with ParticleFlow}
\label{sec:DCTR}
An alternative machine learning method developed to fit parameters is  DCTR~\cite{Andreassen:2019nnm}.
This method is based upon parameterized neural networks~\cite{Baldi:2016fzo} and exploits a relationship between the loss function and the likelihood ratio~\cite{Cranmer:2015bka,Brehmer:2018kdj,Brehmer:2018eca,Brehmer:2018hga,Stoye:2018ovl,Brehmer:2019xox,Andreassen:2019cjw,Erdmann:2020tpv,Hollingsworth:2020kjg,Badiali:2020wal,Andreassen:2020nkr}.

The DCTR method works as follows. 
Suppose we have some parameters $\theta$ and some observables $x$. The probability distribution $p(x|\theta)$ of the observables depends on the values
chosen for $\theta$. An ambitious goal is to learn a function $f(x,\theta)$ which gives the full likelihood distribution of the observables $x$ for any $\theta$ (as in JUNIPR~\cite{Andreassen:2018apy,Andreassen:2019txo}).
In practice, DCTR learns this distribution relative to the distribution over $x$ for a fixed reference value $\theta_0$. To do so, we give it observables ${\bm {x}_{\theta_0}}$ drawn from the distribution at fixed $\theta=\theta_0$ (the reference sample) as well as observables ${\bm{x}_{\theta_S}}$ drawn from the distribution using many values of $\theta \in \theta_S$ (the scanned sample).
We do not tell the network the value of $\theta_0$, however.
Instead, we pretend that $\theta_0$ is equal to $\theta$ and the network will learn that this is inconsistent.
In practice, we train the network over pairs of events $\{x_i, x_i^0 \} \in \{\bm{x}_{\theta_S},\bm{x}_{\theta_0} \}$ chosen over a distribution of $\theta$ values and compute the binary cross-entropy loss
\begin{equation}
    f = \underset{f^{\prime}}{\text{argmin}}\bigg(-\sum_\theta \sum_{ \{x_i, x_i^0 \} \in \{\bm{x}_{\theta_S},\bm{x}_{\theta_0} \},  } \left[ \log\big(f^{\prime}(x_i, \theta)\big) + \log \big(1-f^{\prime}(x_i^0, \theta)\big)  \right]\bigg)
    \label{eqn:dctrclassifier}
\end{equation}
It is important that $f^{\prime}$ in the second term takes $\theta$ and not $\theta_0$, otherwise the classification would be trivial.

Using such a loss function, the DCTR process for inferring model parameters from an ensemble of events involves two steps.
\begin{enumerate}
    \item Train a parameterized classifier $f(x,\theta)$. In the application to top mass extraction, the reference sample has $\theta_0$ corresponding to a fixed mass and tune. The scanned sample has $\theta$ which varies among values of $\mtmc$ and many values for the tune parameters.
    \item Use the function $f(x,\theta)$ for regression. To do so, we re-minimize the loss for an unknown sample compared with an independent sample drawn using the same parameters as the reference sample. Now the network is fixed, but the parameters $\theta$ are varied to minimize the loss. The values which minimize the loss are the prediction.
\end{enumerate}

To give a better sense of how DCTR works, we include a toy  example with a one-dimensional Gaussian in Appendix~\ref{sec:AppDCTR}.
For more details on DCTR, see~\cite{Andreassen:2019nnm} or~\cite{Andreassen:2020gtw}.

\subsection{Network architecture}
\label{sec:PFlow}

\begin{figure}[p]
    \centering
    \includegraphics[width=\linewidth]{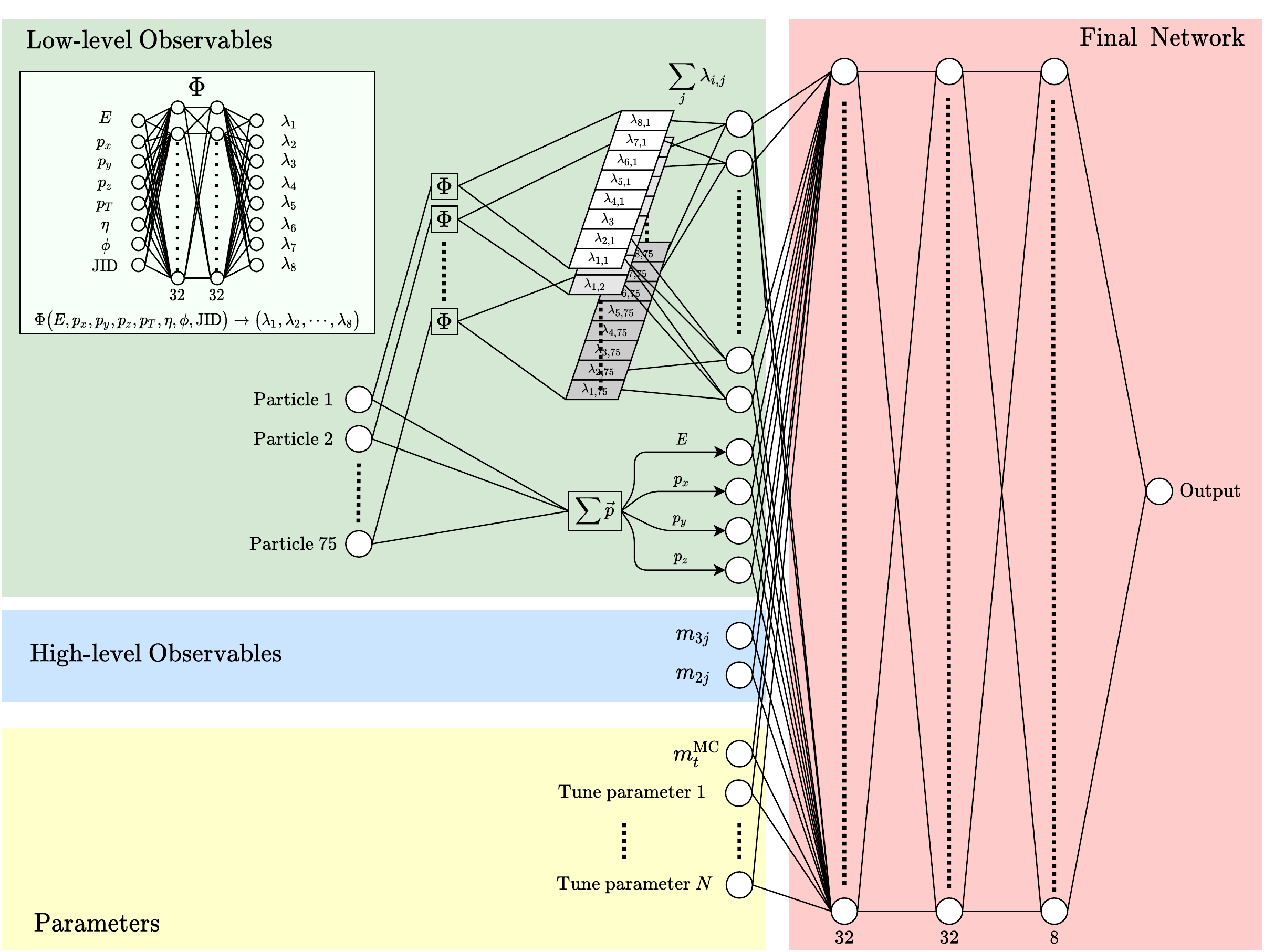}
    \caption{\small{Architecture used to infer the top-quark mass with DCTR.
    The green portion shows the low-level information from the constituents of the jets.
    These are combined with a ParticleFlow Network, denoted by $\Phi$ acting on each constituent, with the resulting output being summed across the particles.
    The $\Phi$ network is shown in the breakout box.
    Next are the high-level inputs of the three-jet and two-jet invariant masses, shown in the blue portion.
    The last elements are the Monte Carlo parameters to be inferred, shown in yellow.
    All of these are combined in the latent space, which is then connected to the final output with a dense neural network shown in red.
    }
    }
    \label{fig:PFlow}
\end{figure}

In order to use DCTR to infer the top-quark mass, we need a parameterized neural network, $f(x,\theta)$ which is flexible enough to learn the likelihood ratio.
The parameter(s) $\theta$ must include $\mtmc$, but can also include the other tune parameters, depending on whether we try to regress those tune parameters or marginalize over them and only extract $\mtmc$.
We find the most effective network takes as input both low-level and high-level observables. The architecture of the network is sketched in Fig.~\ref{fig:PFlow}.

For the high-level variables we take $m_{3j}$ (the ``top mass") and $m_{2j}$ (the ``W mass"), as in previous sections. We consider optionally applying soft-drop jet grooming to the jets before constructing the invariant masses.
These are indicated by the blue portion of the figure.

For low-level observables, following DCTR~\cite{Andreassen:2019nnm} we use four-vectors of the constituent particles of the jets represented with a ParticleFlow Network~\cite{Komiske:2018cqr}.
This is shown in the green region of the figure.
For the inputs to ParticleFlow, we include up to 75 particles, with a maximum of 25 from each of the three jets.
Each particle contains eight input variables: four variables are the four vector in ($p_x, p_y, p_z, E$), three variables are the momenta in a transformed coordinate system ($p_T, \eta, \phi$), and a discrete tag for which jet the constituent came from (0 for the $b$-tagged jet, 1 for the hardest un-tagged jet, and 2 for the softer un-tagged jet).
A function $\Phi$ is applied to each of the particles in the event, mapping from an eight-dimensional input to a $k$-dimensional output.
To ensure that the ordering of the particles is unimportant, each $k$-dimensional output
is symmetrized (summed) over the particles.
These are marked by the $\lambda_k$ nodes in the figure.
For $\Phi$ we use a neural network with two hidden layers.
Each hidden layer contains 32 nodes with the ELU activation function and use a dropout rate of 10\% during training.
The final layer of $\Phi$ contains eight nodes, also using the ELU activation.\footnote{The weights of $\Phi$ are trained along with the rest of the network, but could be pre-trained from a similar application.}
Many applications of ParticleFlow find that a larger latent space is needed, however, we found our results to be much more stable with 8, rather than 16 nodes.
We also tried not including the ParticleFlow part of the network, but found better performance when it is included.
As an additional input, we sum the four-vectors of each of the constituent particles and pass the sum top-quark 4-vector directly to the latent space.

The combined information from the Monte Carlo parameters, the high-level inputs, and the low-level inputs are concatenated together.
From this space, another  neural network is applied to generate the final event level classification, shown in the red region of the figure.
We use three hidden layers with 32, 32, and 8 nodes, respectively.
We again use the ELU activation function and apply a 10\% dropout rate during training.
The output is a single node activated with the sigmoid function.
This results in networks that have between approximately 1400 and 3600 weights, depending on the number of Monte Carlo parameters included in the parameterization.

\subsection{DCTR on a single tune}
\label{sec:DCTRSingleTune}
First we test DCTR's ability to regress $\mtmc$ for a fixed tune (A14 tune 21).
For the fixed reference sample $\theta_0$, we chose the top-quark mass to be 175 GeV.\footnote{We found empirically that the DCTR procedure works better in practice if the Monte Carlo mass of the reference sample is larger than the values to be inferred.}
In the scanned sample $\theta$, we randomly choose $\mtmc$ for each event from a uniform distribution between 170-176 GeV.
We use 1 million events each for the fixed and reference samples.
The data set is split with 25\% for validation and 75\% for training.

As part of the study, we want to see if more information than just the three jet invariant mass can help the network extract the top mass better.
To do so, we allow the network to use only $m_{3j}$; to use $m_{3j}$ and $m_{2j}$; or to use $m_{3j}$, $m_{2j}$ and
the low-level inputs (as described above).
In addition to these observables, the network is also given a value for $\theta =\mtmc$.
For the scanned sample, this is simply the value chosen in the random draw for the event.
In the fixed reference sample, where $\theta_0 = 175$ GeV, the value of $\theta$ input is masked to a random value, chosen from the same range as in the scanned sample.

The networks are trained using the Adam optimizer to minimize the binary cross entropy loss function.
We set the initial learning rate to $10^{-3}$ and use the default $\beta$ values for Adam.
When the loss on the validation set has not improved for 10 epochs, the learning rate is decreased by a factor of $\sqrt{10}$, with a minimum rate of $10^{-6}$.
We also implement early stopping; if the validation loss has not improved for 25 epochs, training is halted.
Training typically takes around 80 epochs.

\begin{figure}[t]
    \centering
    \includegraphics[width=0.8\linewidth]{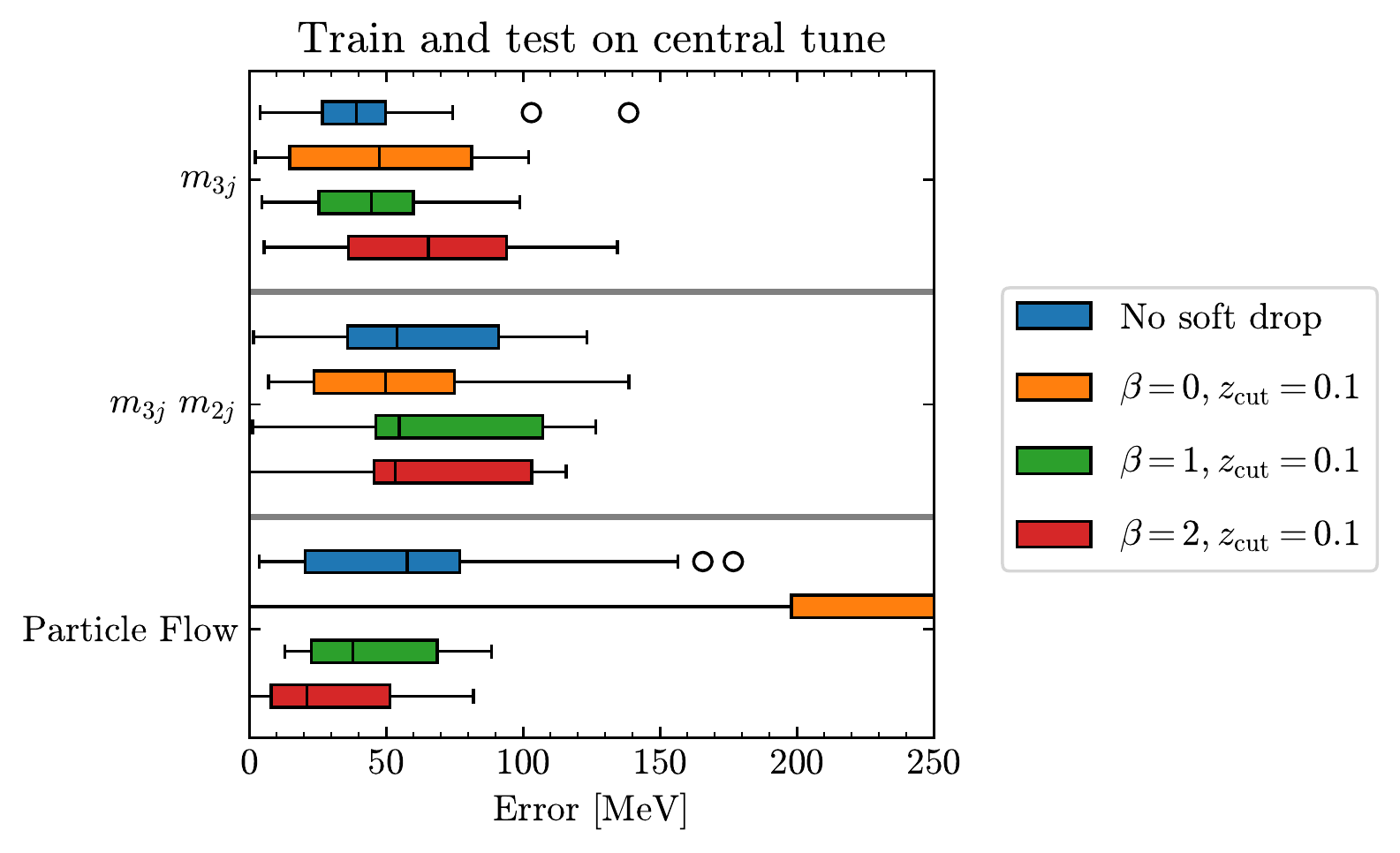}
    \caption{Errors on the mass of the top quark returned by the DCTR methods when trained and tested only on the central tune, for different soft drop parameters.
    Adding extra information (such as the mass of the $W$ jet) does not seem to increase the accuracy.
    The uncertainty for the method is on the order of 50 MeV.
    }
    \label{fig:DCTRCentral}
\end{figure}

After training the networks, the mass is extracted by computing the loss of the classifier between a  test set and an independent reference set.
We repeat this with the same five masses and five iterations of the test sets as in the regression methods presented earlier, with $4\times10^5$ events in each data set.
The loss is minimized for the combined (test and reference) data as a function of $\mtmc$. 

In Fig.~\ref{fig:DCTRCentral} we summarize the results for different soft-drop grooming parameter, and with and without the low-level inputs. 
Numbers shown are the absolute difference between the true and extracted Monte Carlo mass.
The median error for nearly all of the methods here is $\lesssim 50$ MeV, while the maximum error is around 100-150 MeV.
For this exercise, where the tune is fixed and only $\mtmc$ varies, there does not seem to be an advantage to using extra information (such as the mass of the jets from the $W$) or performing jet grooming.
This conclusion will change when we include tune variations.

\subsection{DCTR on Var 1 tunes}
\label{sec:dctr_var1}
We saw that with no tune uncertainties, the DCTR method can regress $\mtmc$ with an uncertainty of order 50 MeV.
When other parameters related to tunes are varied, such as in the showering or hadronization models, DCTR offers multiple ways to proceed. We could train only on a single tune, trying to learn $\mtmc$; we could train on multiple different tunes, again trying to learn only $\mtmc$; or
 we could train over different tuning parameters and try to learn those as well as as $\mtmc$.

To asses which of these options works the best, we train networks on data using the Var1 tunes.
We again use $10^6$ samples for the reference set and $10^6$ sample for the scanned set with 75\% of these samples for training and 25\% for validation.
For the reference set, we use samples drawn from the central tune (tune 21).
The scanned set uses a uniform distribution for the mass ($m_t^{\rm{MC}}$), the color re-connection range, and the strong coupling constant for multiple parton interactions ($\alpha_S^{\rm{MPI}}$).
To remove edge effects, the sampling space is larger than the tune variations we eventually test against.
Explicitly, the ranges are given by
\begin{equation}
\begin{aligned}
    m_t^{\rm{MC}} \in& [170~\rm{GeV}, 176~\rm{GeV}], \\
    {\text{Color re-connection range}}\in& [1.67, 1.75],~\rm{and} \\
    \alpha_S^{\rm{MPI}} \in&[0.116, 0.136],
\end{aligned}
\end{equation}
and there is no correlation in the random samples.
The training procedure is the same as above.

After training the network, we use DCTR to infer the mass (and possibly the color re-connection range and strong coupling) on three different tunes: 21, 23, and 24.
These are the central, up, and down tunes of Var1.
For reference the color re-connection range and the strong coupling constant for the tunes are (1.71, 0.126), (1.73, 0.131), and (1.69, 0.121) for 21, 23, and 24, respectively.
For each test mass, we evaluate the spread in the inferred mass from the different tunes.
This process is repeated for five separate test sets, each with $4\times10^5$ events for the reference and test set.

\begin{figure}[t]
    \centering
    \includegraphics[width=\linewidth]{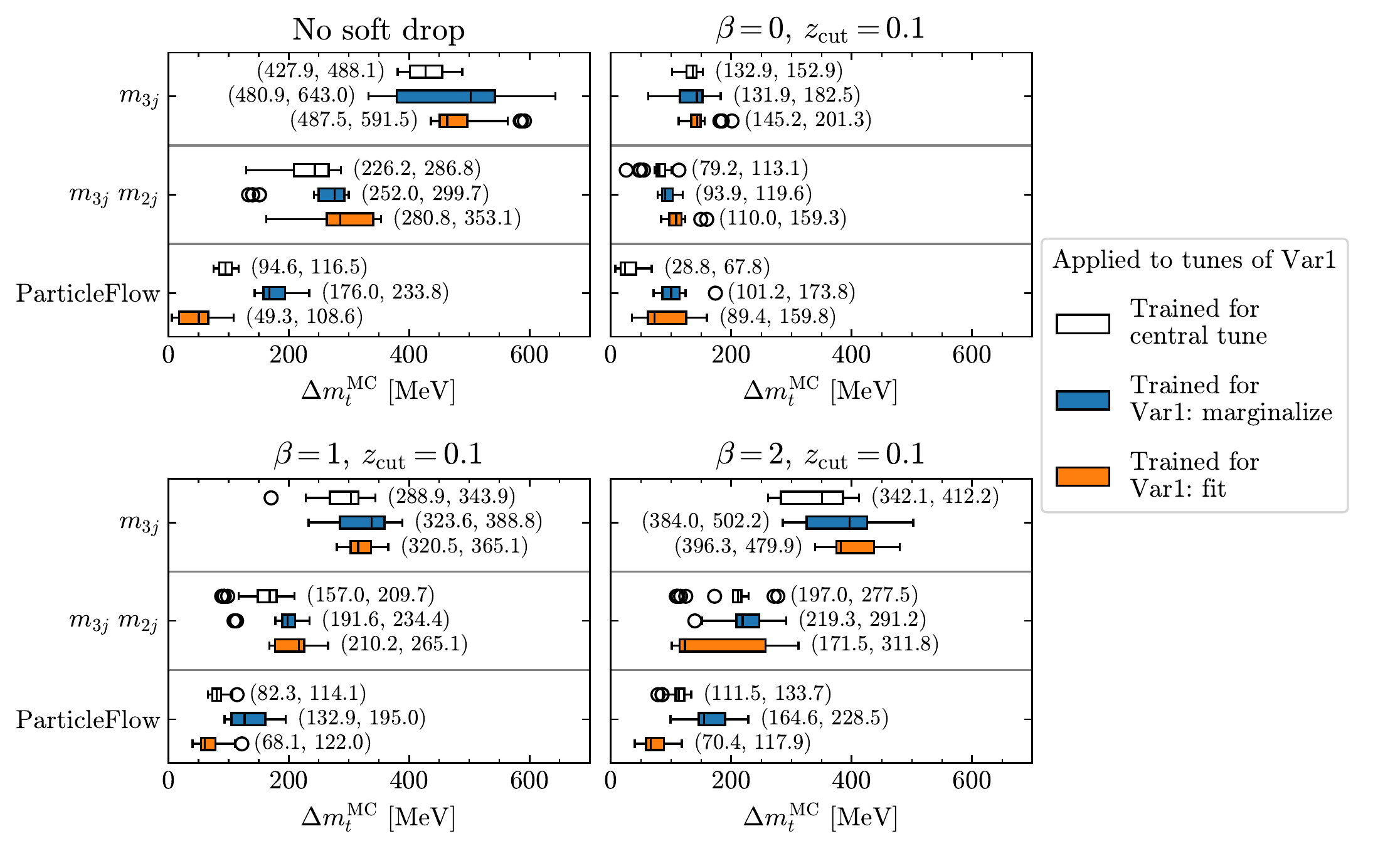}
    \caption{Results testing the input data for DCTR to use as well as the method of inference on VAR1.
    Including both $m_{3j}$ and $m_{2j}$ along with the low-level information captured by ParticleFlow results in the lowest uncertainty.
    DCTR works best when inferring all of the tune parameters (orange) as opposed to marginalizing over them (blue).
    }
    \label{fig:DCTRVar1}
\end{figure}

The results of the spreads are summarized in Fig.~\ref{fig:DCTRVar1} with box-and-whisker plots.
The results for the ungroomed jets are in the upper left panel, using soft drop with $\beta=0$  in the upper right panel, using soft drop with $\beta=1$ in the bottom left, and using soft drop with $\beta=2$ are in the bottom right panel.
In the top row of each panel, the only observable given to the classifier is the three jet invariant mass. The networks of the middle row have access to the two-jet invariant mass in addition, and the bottom row also includes a ParticleFlow network for the constituents of the three jets.

There are many noteworthy trends in these results.
First, we examine how the different grooming methods affect the reconstruction.
We saw before that the using soft drop for the histogram fitting methods greatly reduced the uncertainty.
A similar pattern is observed here, especially when looking at the first two rows (not using ParticleFlow).
For instance, all of the color bars for both the $\beta=0$ and $\beta=1$ panels have significantly lower mean and maximum $\Delta m_t^{\rm{MC}}$ than the corresponding colors for not using soft drop.
The option of soft drop with $\beta=2$ still does better than no soft drop, but not as good as the others.

The next noteworthy trend is that adding more information to the network helps to reduce the uncertainty.
In each panel, the uncertainty is largest when only using $m_{3j}$ and improves when adding in $m_{2j}$.
The uncertainty is further reduced when including the ParticleFlow information in most panels.
However, these networks are more challenging to train and often do not work for the full mass range.
This is why the mean (and median) values drop, while sometimes still having large maximum uncertainties.

The last important observation is that the networks with ParticleFlow do better when they are also fitting to the tune parameters.
The white data is for networks trained on the central tune alone, and thus only capable of inferring the mass.
The blue data sees the scan across the tune parameters, but only tries to infer the mass, while the orange data also infers the tune parameters.
For the $m_{3j}$ alone or $m_{3j}$ and $m_{2j}$ rows, marginalizing over or fitting the tune parameters actually tends to make the uncertainties worse.
With such a small amount of information (either one or two observables), the network does not learn how to correlate the changes in the tune parameters to changes in the observables.
However, when the network also includes ParticleFlow, it can learn these correlations, and thus the uncertainty is reduced when fitting the tune parameters.

To summarize these results, DCTR works better with more input observables.
Using the information contained in the four vectors of the constituents of the jets coming from the top quark decay allows DCTR to correct for differences in the distributions caused by changing the tune parameters.
Therefore, for  the full set of tuning parameters in the next section, we restrict to the case of including ParticleFlow in the inputs and fitting each of the Monte Carlo parameters.

\subsection{DCTR on full set of A14 tunes}
\label{sec:dctr_a14}

\begin{figure}[t]
    \centering
    \includegraphics[width=\linewidth]{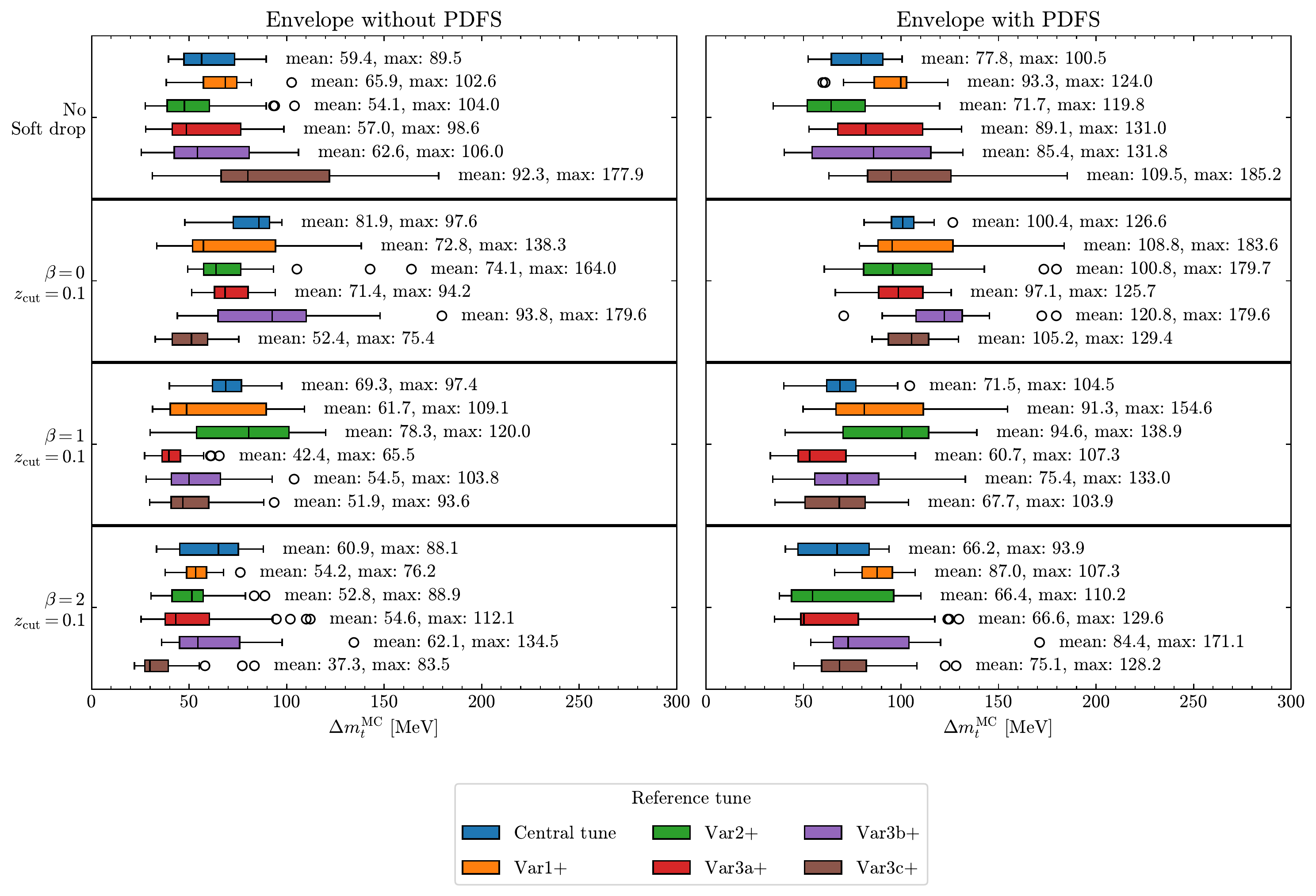}
    \caption{Total correlated uncertainty on the DCTR extraction of $\mtmc$ coming from only the Pythia tunes (left) and additionally including the PDF variation (right).
    The different rows show various grooming options and the colors denote different tunes used as the reference sample $(\bm{\theta_0})$ for the DCTR training.
    The no soft drop option trained using the central tune is chosen to compare with the other regression methods.
    }
    \label{Fig:DCTRA14}
\end{figure}

We now apply the DCTR methodology to the A14 variations.
The PDF variations are not included in the training, although we do evaluate on them.
The reason for this is that PDF selection is a discrete choice, and DCTR is designed to work on continuous parameters. That is, there are not specific Monte Carlo parameters for DCTR to infer from the different PDFs. One can nevertheless assume that PDF variations are within the range of other tune variations and test how well DCTR works on samples generated with different PDF sets. 

When considering multiple tunes, we must also decide which
tune to use as the reference sample $\theta_0$.
Using the central tune would be the most obvious choice.
However, since we found that the DCTR algorithm works better when we use higher reference masses, we allow for the possibility that it will work better using non-central tunes. 
We therefore test taking as the reference sample both the central tune as well as each of the ``+'' tunes for each of the variations.

For the scanned sample, the Monte Carlo parameters are randomly sampled for each event.
The mass is drawn from a uniform distribution with a range of 170 GeV to 176 GeV.
The tune parameters are sampled from $(\min z - 0.5 \Delta z, \max z + 0.5 \Delta z)$, where $z$ represents the value of and individual tune parameter, $\min z$ is the minimum value across the variations, $\max z$ is the maximum value across the variations, and $\Delta z$ is the difference between the maximum and the minimum.
The sampling space is larger than the values we will be testing at to remove possible edge effects.

The results are summarized in Fig.~\ref{Fig:DCTRA14}.
The left panel shows the total envelope of $\Delta \mtmc$ caused by changing the Pythia tune parameters across the 11 variations.
The right panel additionally includes the 3 remaining PDF variations.
The different rows show different amounts of grooming, with no grooming on top and the three different soft drop options in the remaining rows.
Each color denotes a different tune used as the reference set.

Overall, there is not an obvious best choice for the reference tune.
In some choices of grooming, one tune will do better, but then it will not do as well on the different grooming choice.
Similarly, some reference tunes that do well without the PDF variations do not generalize as well to including the variations from the PDF.
However, we do note that using soft drop with $\beta = 0$ seems to consistently lead to worse results.

With an unclear best option, we chose to use no soft drop trained on the central tune to compare with the other regression models.
This option generalizes well from training without the PDFs to including the PDF variations, only increasing the uncertainty by around 10 MeV.

\begin{figure}[p]
    \centering
    \includegraphics[width=0.95\linewidth]{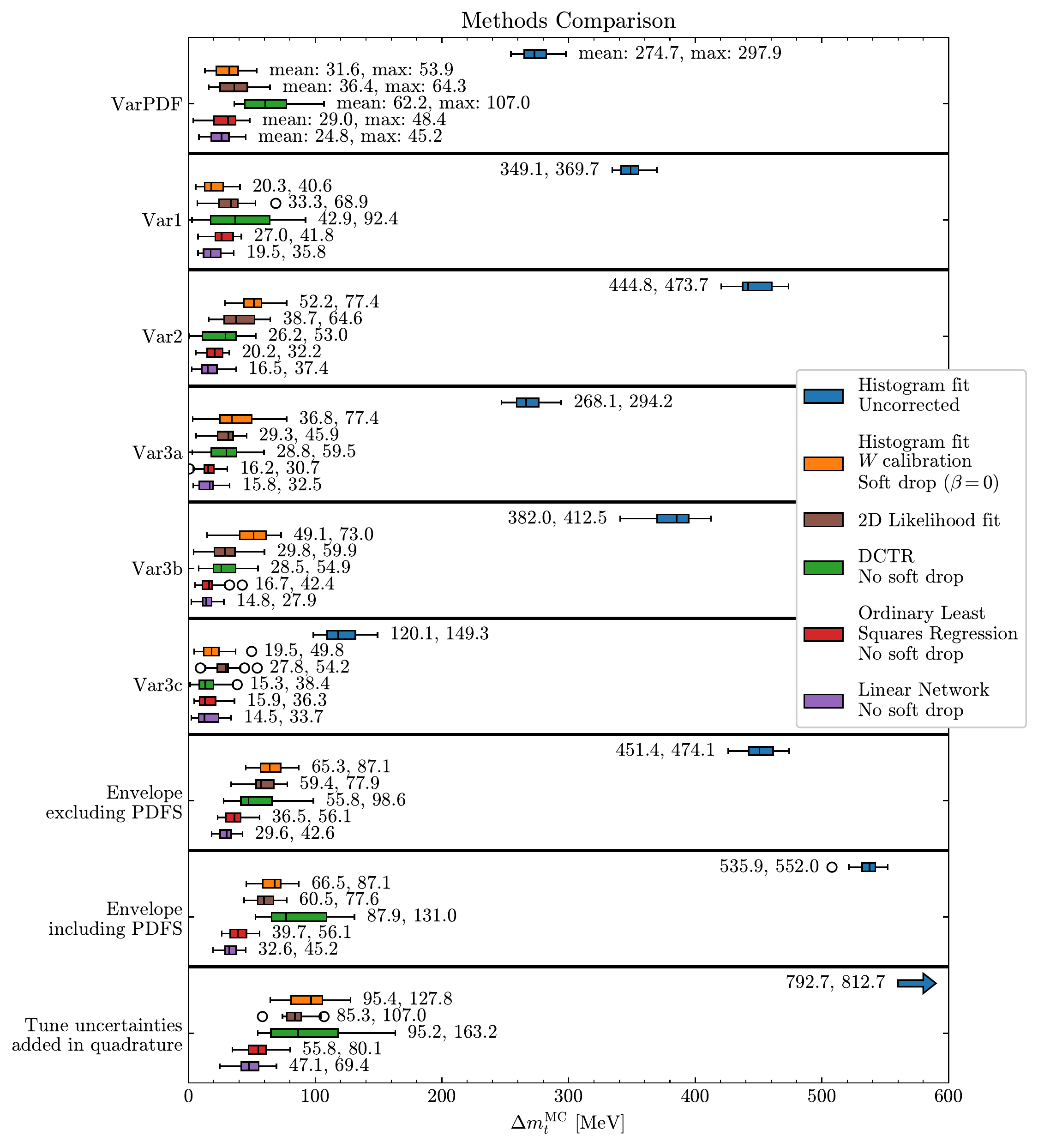}
    \caption{Summary comparison of the different methods. Upper six boxes show top quark tuning uncertainty among six different families of variations. The bottom boxes combine the separate uncertainties either by taking the envelope over the variations or by adding the uncertainties in quadrature. Here, DCTR uses ParticleFlow to simultaneously extract the mass and the tune parameters, and both the ordinary least squares regression and linear network are trained on ensembles of re-sampled events. The various methods are discussed more in the text. }
    \label{fig:AllMethods}
\end{figure}

\section{Conclusions}
\label{sec:discussion}
In this paper we have investigated an example of how machine-learning methods could help with measurement tasks at particle colliders. In particular, we explored a situation in which regression is assisted by learning simultaneously on an ensemble of  events rather than on individual events. Despite the fact the individual events are totally uncorrelated, we find the best performance when variables constructed from  the events are concatenated into an array, sorted, then input to the regression algorithm. 

The case we explored is when the measurement is done by curve-fitting to simulated data to regress a single simulation parameter marginalizing over other parameters. In particular, we looked at the top-quark mass measurement. The traditional method is to extract the top-quark Monte Carlo mass $\mtmc$ by fitting to histograms, and then to estimate an uncertainty $\Delta \mtmc$ on this extracted value due to Monte-Carlo tuning uncertainty. This tuning uncertainty might be of order 500 MeV, which is comparable to statistical uncertainties, experimental systematic uncertainties, and theoretical uncertainties (such as converting the top $\mtmc$ mass to a short-distance mass). In this paper we focus only on reducing the tuning uncertainty.

We explored 4 classes of methodology to regress $\mtmc$. First, we looked at histogram curve fitting. Uncertainties from this method are around 500 MeV but reduce to around 100 MeV if jet substructure techniques are used to clean the data before fitting (as shown in~\cite{Andreassen:2017ugs}). Second, we used a 2D likelihood fit using the raw $m_{2j}$ and $m_{3j}$ observables incorporating a nuisance parameter to account for Monte Carlo tune differences. Third, we looked at linear regression techniques, both using a dense but shallow linear network and using ordinary least squares regression. Fourth, we used a machine-learning method called DCTR. DCTR is a two-step method: first the weights of a distribution are learned as a function of tuning parameters relative to a fiducial sample, and second the tuning parameters are optimized for a given test data sample. 

The results of our study are summarized in detail in  Fig.~\ref{fig:AllMethods}, with more details of each method in the appropriate section, and fewer details in the concise summary plot shown in  Fig.~\ref{fig:ErrorSummary}. 
Fig.~\ref{fig:AllMethods} shows the box-and-whisker plots for the different families of variations, while the final three rows show different methods of combining the variations. Probably the most realistic estimate of error is the ``envelope including PDFs", which means we take the maximum and minimum values for the fit top mass across the  A14 tunes. Such an approach assumes that the tunes are correlated and that actual data will lie somewhere within the complete range of variation. 
For completeness, we also include a more conservative estimate where each variation is assumed to probe completely different physics and is uncorrelated with other tune variations. In this case we add the errors in quadrature.
The PDF variations are special because they are discrete: there is no way to interpolate between different A14 PDF sets as we could for other parameters such as $\alpha_s$. \footnote{There are other ways to study continuous variations of the PDF sets, but they are beyond the scope of this study.} It is unclear how to train DCTR for PDF variations because of this complication. We thus include also numbers for the total envelope not including PDFs. Note that for most tune parameters there is no ``right" answer: approximations such as the parton shower are made so the data can never be described perfectly. Thus in the context of a particular Monte Carlo simulation there is an irreducible uncertainty on how well the data could ever be described. In contrast, there is, in principle, a right answer for the PDFs, although in practice they are always used and fit in conjunction with calculations at a fixed perturbative order. In any case, our purpose is not advocating any particular choice of how to combine errors for an experimental analysis. We are simply providing various metrics by which the different methodologies can be compared.

The main take-home lesson from the summary in Fig.~\ref{fig:AllMethods} is that in practice regression on sorted event ensembles does better than the classical histogram fitting approach or the profile likelihood fit.
Using such methods reduce the Monte-Carlo-based uncertainty on current extractions on the top-quark mass from LHC data, perhaps even by a factor of 2.
This is on top of the reduction gained from using jet grooming as advocated in~\cite{Andreassen:2017ugs}. We found that DCTR works fine for the families of variations for which it was trained (not the PDFs), and has similar uncertainties to the histogram method. It has the potential benefit of being able to fit other tuning parameters well, but if one is only interested in a specific measurement, such as the top mass, then DCTR may be over-kill. Indeed, DCTR is somewhat challenging to implement and train, sensitive to how the inputs are refined, and requires some hyperparameter adjustment to get reasonable results at all. Its killer application may be more along the lines of~\cite{Andreassen:2020gtw} than direct parameter estimation. 

Having established that a regression on sorted event ensembles is more effective than curve fitting a histogram, we also looked into what features of the ensemble the regression uses. 
In contrast to the histogram fitting approach, which focuses on the center of each observable's distribution, ensemble learning methods can weigh various parts of each observable's distribution  differently. In Section~\ref{sec:dense}, we showed how the ensemble methods can use the center of the $m_{3j}$ distribution to learn the difference in masses, while using the upper tail of $m_{3j}$ and the other observables to correct for the difference in tunes.

It is worth emphasizing that the point of this study is not a total numerical estimate for the uncertainty. Values throughout this paper do not include any estimate of experimental systematic effects on the Monte Carlo tuning uncertainties, such as smearing due to jet energy resolution or detector effects. Thus one should not take the absolute size of the numbers as indicative that the tuning uncertainty could be reduced to the 30 MeV level. We do, however, conclude that linear regression, either through a linear network or an ordinary least squared regression on an ensemble of events, is a promising technique that has the potential to significantly reduce the dependence of the measured top-quark mass on Monte Carlo tuning parameters beyond the methodology already being employed.
Although the uncertainty from marginalizing over unphysical parameters in simulation is smaller than other sources, it can be important for precision studies such as SUSY searches or vacuum stability, as discussed in the introduction.

In conclusion, we have shown that machine learning regression methods can work most effectively when trained on a sorted ensemble of uncorrelated events. We found these methods can improve upon a traditional histogram-fitting procedure for determining the top-quark Monte Carlo mass. In particular, performing linear regression using a shallow but dense network trained on sorted ensembles of events (30,000 at a time in our study) seems to combine excellent performance with simplicity.

\section*{Acknowledgements}
We thank Anders Andreassen and Benjamin Nachman for feedback on this manuscript.
The computations in this paper were run on the FASRC Cannon cluster supported by the FAS Division of Science Research Computing Group at Harvard University.
KF is supported by the National Science Foundation Graduate Research Fellowship Program under Grant No. DGE1745303.
BO was supported in part by the U.S. Department of Energy (DOE) under contract DE-SC0013607 and DOE Grant No. DE-SC0020223.
MS was supported by DOE Grant No. DE-SC0013607.
This work is supported by the National Science Foundation under Cooperative Agreement PHY-2019786 (The NSF AI Institute for Artificial Intelligence and Fundamental Interactions, http://iaifi.org/)

\appendix
\section{DCTR on a toy model}
\label{sec:AppDCTR}

The overall idea of the DCTR method to extract the top mass is inspired by finding the the value of the Monte Carlo mass (and other tune parameters) which are most likely to have produced the data.
In order to assess this, one needs the likelihood function covering the range of data and parameters.
However, this is extremely difficult to obtain.
Instead, DCTR uses the fact that, given two data sets, the likelihood ratio between the data sets is given by an ideal classifier.
While we cannot access the likelihood function itself, it is still possible to find the parameters which maximize the likelihood using the ratio.
We gain access to an approximation of the likelihood ratio using a well-trained, flexible, neural network, which is close to an ideal classifier.
In this appendix, we review the two of the main components necessary for DCTR to work, (a) training a parameterized neural network to find the likelihood ratio and (b) using the likelihood ratio to maximize the likelihood and infer the most probable Monte Carlo parameter.

In a parameterized neural network, unobserved properties are included as input to the network.
This can be useful when needing to scan over a property.
For instance, when looking for BSM physics, the mass of a new resonance is unknown, and a classifier trained at one mass will be sub-optimal if the mass is substantially different.
Rather than training many classifiers for different masses, one can train a single classifier where the mass is included as an input.
This helps the classifier to interpolate between masses and reduces the amount of training data needed, because features are shared across the feature space.

The first step is to use a parameterized neural network as a classifier to derive an estimate for the likelihood ratio between samples. In our full set up, we include the Monte Carlo mass of the top quark as an input parameter. For this appendix, we will start by considering a simpler example. Consider the case of Gaussian distributions with different values for the mean $\mu$. Draw samples $x$ from this Gaussian, where the mean $\mu$ is changed for each draw.
This produces a two dimensional array $S = (\mu, x)$. 
We can train a network $f(\mu, x)$ to distinguish between $S$ and a uniform 2D distribution, $U$. This network will yield the ratio of their probability densities at any given point,
\begin{equation}
    f(\mu, x) = \frac{S(\mu, x)}{S(\mu, x) + U(\mu, x)}~.
\end{equation}
The probability density for the data distribution can easily be solved for in terms of the output of the network,
\begin{equation}
    S(\mu, x) = \frac{f(\mu, x)}{1 - f(\mu, x)} ~ U(\mu, x)~.
\end{equation}
From this, we can use $f$ to obtain the probability density of $S$ for any $\mu$ and $x$ within the range of the training data.
This is possible because we know the probability density of the uniform distribution which we were using as a reference.
In fact, we did not need to use a uniform distribution at all; the processes generalizes to choosing a different fixed reference sample. For the rest of this simple example, we will use a Gaussian with fixed mean $\mu=3$ as the reference sample. In the main text, we use a sample of events with a fixed Monte Carlo top-quark mass and fixed tune as a reference, since a uniform distribution does not make sense in the context of jets with differing numbers of particles.

\begin{figure}[t]
    \centering
    \includegraphics[width=0.9\linewidth]{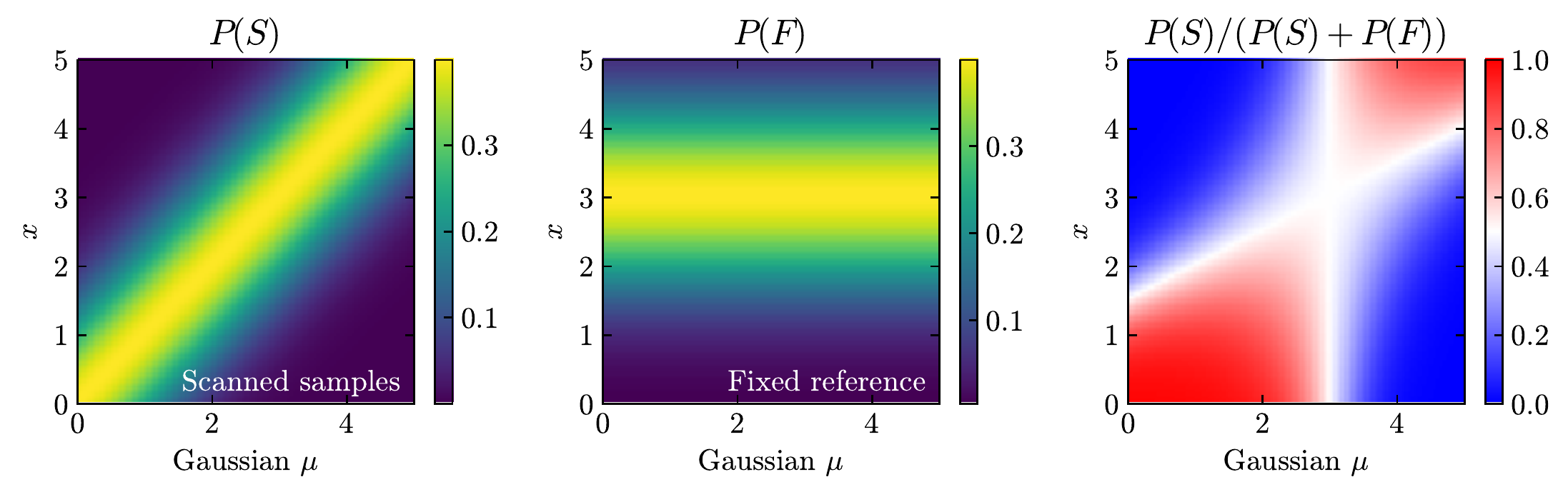}
    \caption{DCTR works by using a classifier to approximate the likelihood ratio.
    In this example, we use Gaussian distributions to show the procedure using exact likelihood ratios.
    The left panel shows the probability density for a sample of data in which the mean of the Gaussian can take any value between 0 and 5; this is referred to as the scanned sample.
    The middle panel shows the probability density for the fixed sample, where the mean is always set to 3 and $\mu$ has no meaning.
    The right panel shows what the probability is for a point in the $\mu, x$ parameter space to come from the scanned sample as opposed to the fixed sample, which is the result of an ideal classifier.
    }
    \label{fig:DCTRGaussianSetup}
\end{figure}

This is shown in Fig.~\ref{fig:DCTRGaussianSetup}. Specifically, we show this initial setup and the idealized network output.
The first panel shows the probability density for our scanned sample $P(S)$, where the $x$-axis denotes the Monte Carlo parameter and the $y$ axis denotes the observed value.
The middle panel displays the same probability density for the fixed reference sample, $P(F)$. We can see that P(F) is uniform for all values along the $x$-axis (the fake input parameters) but has a non-uniform $y$-axis, since the sample is drawn with a specific fixed parameter.
The final panel shows the ratio of these probability densities,
\begin{equation}
    R(\mu, x) = \frac{P\big(S(\mu, x)\big)}{P\big(S(\mu, x)\big) + P\big(F(x)\big)},
\end{equation}
which would be the output of an ideal network trained to classify events as coming from the scanned or reference sets.

\begin{figure}[t]
    \centering
    \includegraphics[width=\linewidth]{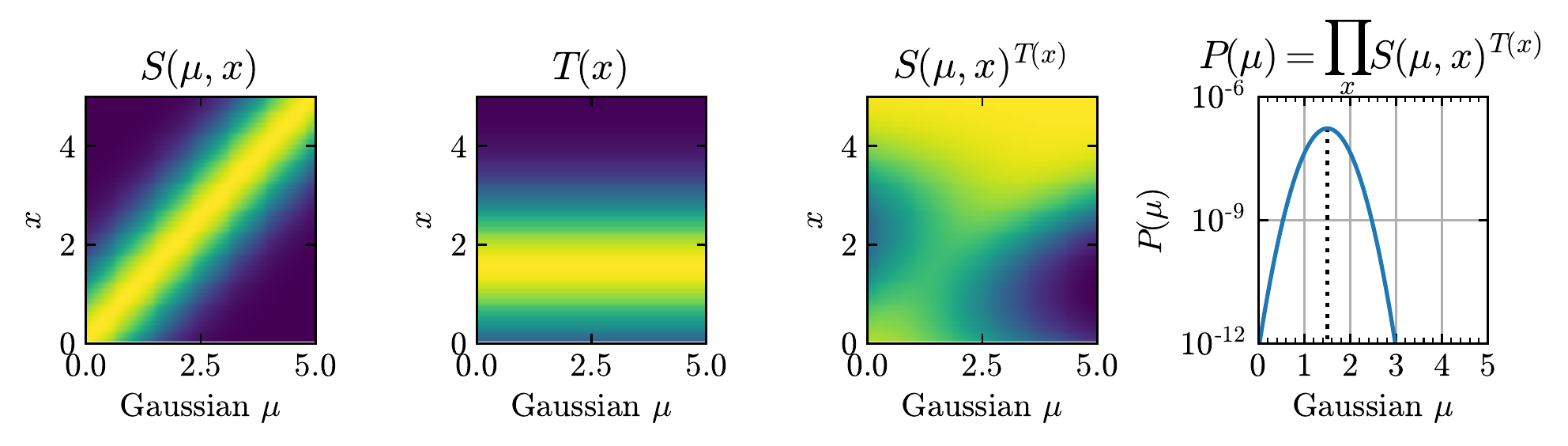}
    \caption{If the full probability density is known (shown in the left panel), the parameter of an unknown sample (second panel) can be obtained by maximizing the likelihood. The third panel shows the probability of observing each element in the unknown set, $T$.
    The final panel shown the total probability, obtained by taking the product of the individual probabilities, and is maximized at the true underlying parameter.
    }
    \label{fig:MaxLikelihood}
\end{figure}

The next step is to use the classifier to infer the most probable Monte Carlo parameter for a new dataset.
In this example, we use $\mu=1.5$ as the new dataset, and denote this set by $T(x)$.
If we had access to the full likelihood $S(\mu, x)$, we could infer the value of $\mu$ by multiplying the probabilities from each event in $T$ to maximize the likelihood,
\begin{equation}
    \hat{\mu}   = \underset{\mu}{\text{argmax}} \prod_{x_i \in T} S(\mu, x_i)
                = \underset{\mu}{\text{argmax}} \prod_{x} S(\mu, x)^{T(x)}.
\end{equation}
In going from the first expression to the second expression, we transition from discrete to continuous distributions. An example of this is shown in Fig.~\ref{fig:MaxLikelihood}.
The first panel again shows the true probability density $S(\mu, x)$ and the second panel shows the distribution of $x$ for the unknown set.
The third panel shows $S(\mu, x)^{T(x)}$ which is the the probability of observing each element in $T$ given $S$.
In the final panel, we take the product of the probabilities to obtain the total probability of obtaining the data $T$ as a function of $\mu$.
The true value is the most probable one.

When using DCTR, we do not actually have access to $S(\mu, x)$ but only the ratio $R(\mu, x)$. However, a similar procedure still works. We want to maximize the probability, but now we must also include the reference set.
This is done by maximizing the likelihood that the new set will be classified as part of the scanned set while the reference set will be classified as the reference set.
We define the total probability of events from $T(x)$ to get classified as coming from the scanned sample as
\begin{equation}
    P_T(\mu) = \prod_x R(\mu, x)^{T(x)}.
    \label{eq:ptmu}
\end{equation}
Similarly, let $P_F(\mu)$ define the probability of events from the fixed sample $F(x)$ getting classified correctly,
\begin{equation}
    P_F(\mu) = \prod_x \big(1 - R(\mu, x)\big)^{F(x)}~.
    \label{eq:pfmu}
\end{equation}
Combining these two expressions yields total probability of classification,
\begin{equation}
    C(\mu) = \prod_x R(\mu, x)^{T(x)} \big(1 - R(\mu, x)\big)^{F(x)}.
    \label{eq:cmu}
\end{equation}
The value of $\mu$ which maximizes $C(\mu)$ then corresponds to the most probable value to have produced the test dataset, $T$.

\begin{figure}[t]
    \centering
    \includegraphics[width=0.81\linewidth]{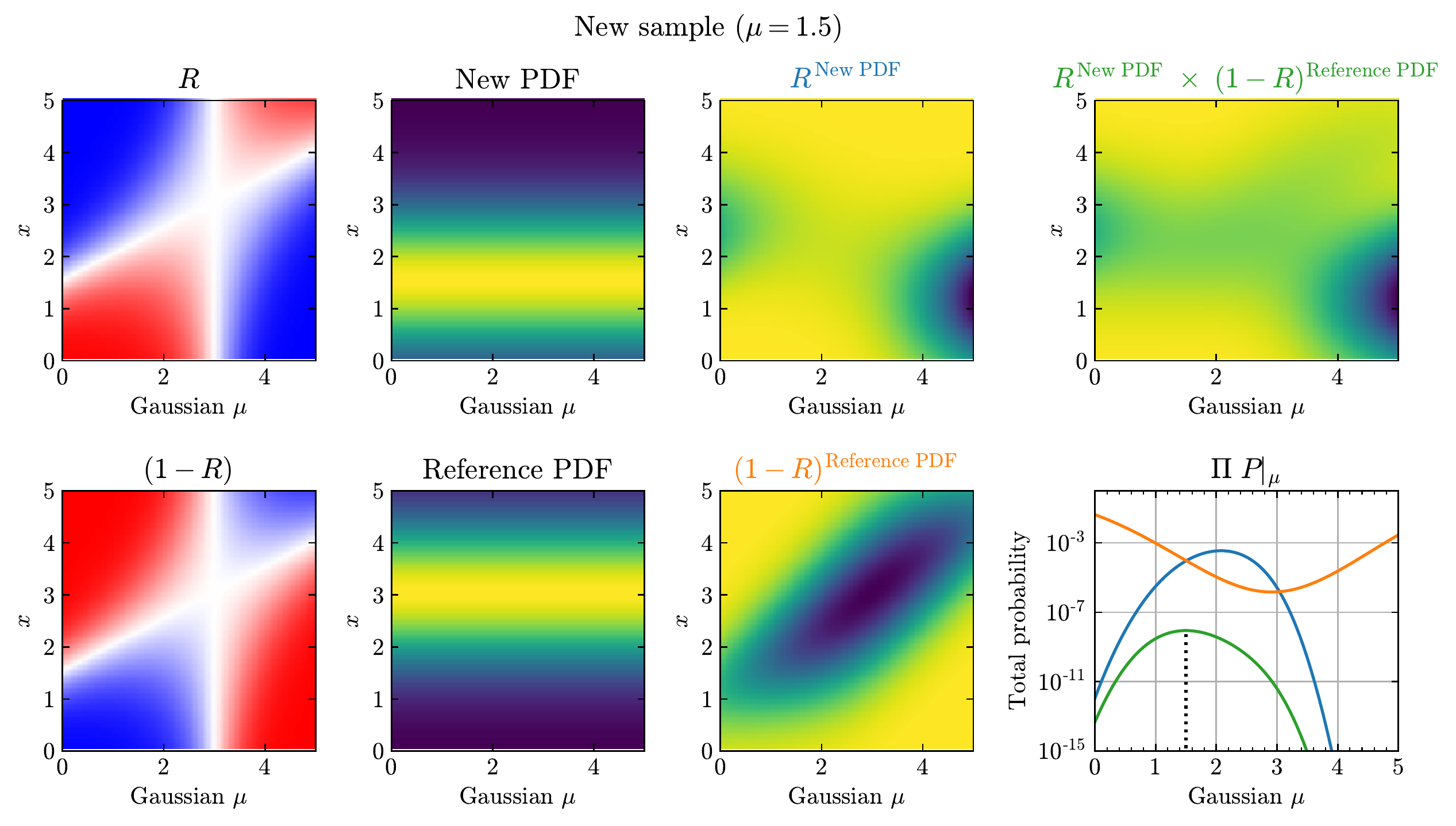}
    \caption{The classifier is now applied to a new unknown sample and the reference sample.
    The upper left panel shows the classifier output $R$, which is the probability for a point in the space to belong the the scanned sample.
    Similarly, $(1-R)$ is the probability to have come from the fixed sample, and is shown in the lower left panel.
    The probability density for the new set and the reference set are displayed in the second column, where neither depends on the model parameter $\mu$.
    The third column displays the classifier output (R or 1-R) convolved with the probability distributions.
    The total probability is displayed in the top right panel; it is the product of the individual probabilities for each sample in the datasets.
    In the bottom right panel, the product of the individual probabilities along the $y$ axis is shown as a function of the model parameter.
    The blue, orange, and green lines denote the probabilities for the new sample, the reference sample, and the combination, respectively.
    The green line is maximized at the value of the unknown parameter used to generate the new PDF.
    }
    \label{fig:DCTRGaussianExample}
\end{figure}

We make this more explicit in Fig.~\ref{fig:DCTRGaussianExample}.
The panels on the left show the output of the ideal classifier (the ratio of probability densities from the scanned and fixed sample) for the scanned (top) and reference (bottom) datasets.
The second column shows the new test PDF which we are trying to infer (top) and the reference PDF (bottom).
The third column displays the classifier output (R or 1-R) convolved with the probability distributions.
The top panel in the last column displays the product of these.
In the bottom right panel, we show the total probabilities.
The new set $P_T(\mu)$ is shown in blue, the fixed reference set $P_F(\mu)$ is shown in orange, and $C(\mu)$ is the green line.
Note that the blue and orange lines have quite different shapes, however, when they are combined to make the green line, it is maximized at $\mu=1.5$, which is the value of the test set.
The works for all values of $\mu$; a video showing a scan can be found at \url{https://bostdiek.github.io/Videos/DCTR_Gaussian_Example.mp4}.

While DCTR is overly complicated for a single dimension, it can prove useful when the datasets have many dimensions.
We now generalize the method by taking the single observable $x$ to be set of observations, $x_i\rightarrow \mathbf{X}_i$, where the subscript represents a given event.
Similarly, the underlying parameter $\mu$ is generalized to many model parameters $\mu \rightarrow {\bm{\theta}}$.
In many dimension, an explicit likelihood ratio can be challenging to obtain, thus a neural network will be used as an approximation.
The network is trained to classify events from a scanned set $x_i\in \bm{x_{\theta_S}}$ from events in a fixed reference set $x_i \in \bm{x_{\theta_0}}$.
We now represent the network output by $f(x, \theta)$ and train it to maximize the probability of correctly assigning the training events.
Namely,
\begin{equation}
    \begin{aligned}
    f(x, \theta) &= \underset{f^{\prime}}{\text{argmax}} \bigg(\prod_{x_i \in \bm{x_{\theta_S}}} f^{\prime}(x_i, \theta) \times \prod_{x_i \in \bm{x_{\theta_0}}} \big(1 - f^{\prime}(x_i, \theta)\big) \bigg) \\
     &= \underset{f^{\prime}}{\text{argmin}} \bigg( - \sum_{x_i \in \bm{x_{\theta_S}}} \log f^{\prime}(x_i, \theta) - \sum_{x_i \in \bm{x_{\theta_0}}} \log\big(1 - f^{\prime}(x_i, \theta)\big) \bigg)
    \end{aligned}.
\end{equation}
The second line is just the usual binary cross-entropy loss function which is used to train binary classification neural networks.

Once the network is trained, we can infer the parameters of a new data, $x_i \in \bm{x_{\theta_T}}$ set by minimizing the loss of classifying the new data versus the reference set.
Thus,
\begin{equation}
    \begin{aligned}
    \hat{\theta} &= \underset{\theta^{\prime}}{\text{argmin}} \bigg( - \sum_{x_i \in \bm{x_{\theta_T}}} \log f^{\prime}(x_i, \theta^{\prime}) - \sum_{x_i \in \bm{x_{\theta_0}}} \log\big(1 - f(x_i, \theta^{\prime})\big) \bigg) \\
    \end{aligned}.
\end{equation}
This is equivalent to maximizing the probability as we did in the Gaussian example.

\section{Training Curves}
\label{sec:TrainingCurves}

\begin{figure}[t]
     \centering
     \begin{subfigure}[b]{0.45\textwidth}
         \centering
         \includegraphics[width=\textwidth]{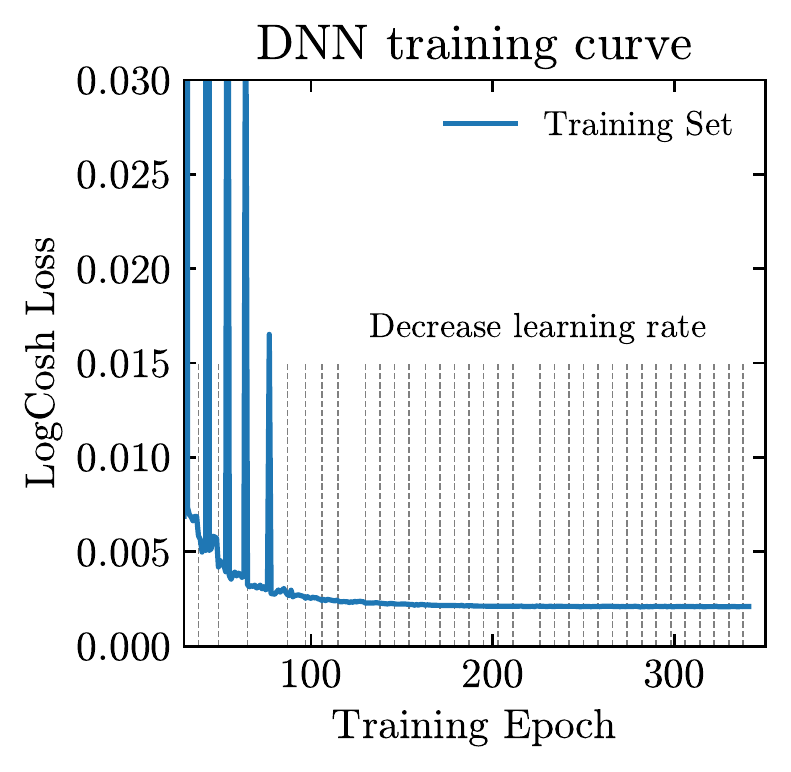}
        \captionsetup{justification=raggedleft, singlelinecheck=false}%
        \caption{\phantom{SSSSSSSSSsspace}}     
     \end{subfigure}
     \hfill
     \begin{subfigure}[b]{0.45\textwidth}
         \centering         \includegraphics[width=\textwidth]{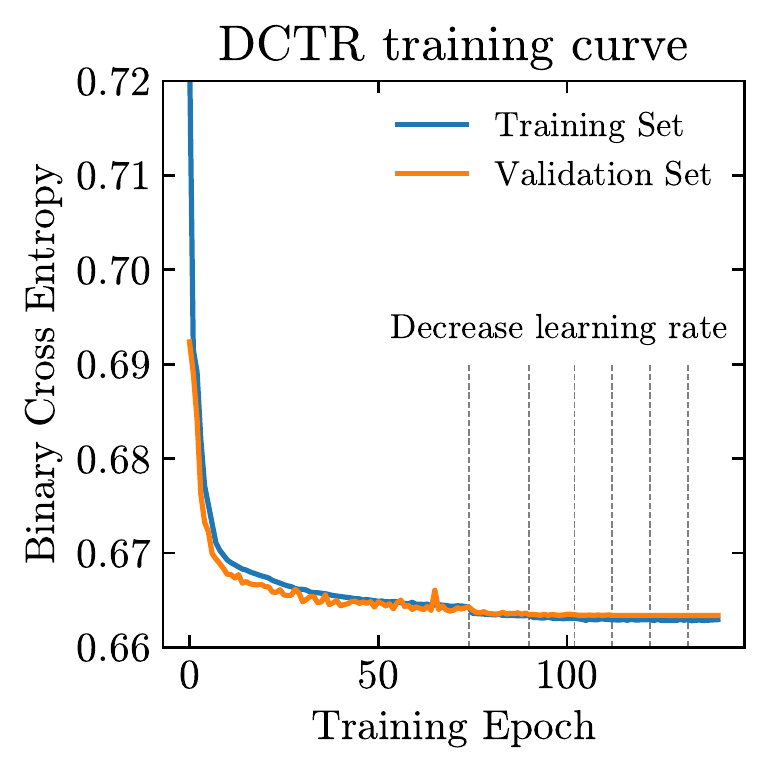}
        \captionsetup{justification=raggedleft, singlelinecheck=false}%
        \caption{\phantom{SSSSSSSSSsspace}}
     \end{subfigure}
        \caption{(a): Training loss curve for linear network. (b): Training loss curve for DCTR classifier.}
        \label{fig:loss curves}
\end{figure}

For completeness, we show training loss curves for DCTR and the linear network. As can be seen, training is stable and early stopping only becomes relevant once a plateau has been reached.

\bibliographystyle{utphys}
\bibliography{Bibfile}

\end{document}